\documentclass[12pt,a4paper,english]{article}
  \pdfoutput=1
\usepackage{lmodern}
\usepackage{hyperref}

\usepackage{ulem}
\usepackage[utf8]{inputenc}
\usepackage{geometry}
\usepackage{color}
\usepackage{float}
\usepackage{amsmath}
\usepackage{amssymb} 
\usepackage{graphicx}
\usepackage{setspace}
\usepackage{subfigure}
\usepackage{amsthm}
\usepackage{amsfonts}
\usepackage{braket}

\makeatletter

\usepackage{slashed}
\usepackage{cite}
\date{} 

\makeatother

\usepackage{babel}

\begin{document}

\title{Holographic Subregion Complexity in
Einstein-Born-Infeld theory}

\author{Yi Ling$^{1,2}$,\thanks{lingy@ihep.ac.cn, co-first author} 
Yuxuan Liu$^{1,2}$,\thanks{liuyuxuan@ihep.ac.cn, co-first author, correspondence author}
Cheng-Yong Zhang$^{3}$, \thanks{zhangchengyong@fudan.edu.cn}}
\maketitle
\begin{center}
\textsl{$^{1}$Institute of High Energy Physics, Chinese Academy of Sciences, Beijing 100049, China}\\
\textsl{$^{2}$School of Physics, University of Chinese Academy of Sciences,
Beijing 100049, China}\\
\textsl{$^{3}$Department of Physics and Center for Field Theory and Particle Physics, Fudan University, Shanghai 200433, China}
\par\end{center}
\begin{abstract}

We numerically investigate the evolution of the holographic
subregion complexity during a quench process in
Einstein-Born-Infeld theory. Based on the subregion CV conjecture,
we argue that the subregion complexity can be treated as a probe
to explore the interior of the black hole. The effects of the
nonlinear parameter and the charge on the evolution of the
holographic subregion complexity are also investigated. When the
charge is sufficiently large, it not only changes the evolution
pattern of the subregion complexity, but also washes out the second stage featured by linear growth.

\end{abstract}

\newpage
\tableofcontents{}

\section{Introduction}

The holographic nature of spacetime can be manifestly disclosed
by AdS/CFT correspondence \cite{Maldacena:1997re,Gubser:1998bc,Witten:1998qj}. Recently it
has been proposed that the exponential growth of the interior of a black hole can be described by a quantity in the quantum field
theory on the boundary \cite{Stanford:2014jda}. Specifically, it
has been conjectured that the quantum computational complexity is
equal to the volume of Einstein-Rosen Bridge (ERB) (CV conjecture)
\cite{Stanford:2014jda}. An outstanding model has been considered
in the AdS-Schwarzschild geometry, where the maximal volume of
codimension-one surface $\Sigma$ bounded by the boundary time
$t_{L}$ and $t_{R}$ (where $L$ and $R$ label the left and the
right boundary, respectively.) is dual to the quantum
computational complexity of a boundary state
$\ket{TFD(t_{L},t_{R})}$ relative to the reference state
$\ket{TFD}$:
\begin{align}
  &C(t_{L},t_{R})=\frac{V_{\Sigma}(t_{L},t_{R})}{G_{N}l},\label{CV} \\
  &\ket{TFD(t_{L},t_{R})} =U(t_{L})U(t_{R})\ket{TFD}\nonumber.
\end{align}
where $U(t_{L})$, $U(t_{R})$ are quantum gates with $U$ labelling
the time evolution operator in the boundary theory. $G_N$ is the
gravitational constant and $l$ is a certain length scale.

Because of the ambiguity of choosing the radius $l$, it
has been further conjectured that the quantum computational
complexity is equal to the gravitational action on the Wheeler
DeWitt (WDW) patch (CA conjecture) \cite{Brown:2015bva}:
\begin{equation}
  C(t_L,t_R)=\frac{A_{WDW}(t_L,t_R)}{\pi\hbar}.
\end{equation}
The WDW patch is the domain of dependence of a cauchy slice anchored at some boundary time $t_L$ and $t_R$.

These two conjectures have been extensively testified in
literature. On the gravity side, the growth behavior of the action
as well as the maximal volume has firstly been investigated in the
late time limit \cite{Zhang:2017nth,Cano:2018aqi,Couch:2017yil,Auzzi:2018zdu}, and then for
the full-time period
\cite{HosseiniMansoori:2018gdu,Ghodrati:2018hss,Mahapatra:2018gig,Alishahiha:2018tep,An:2018xhv,Swingle:2017zcd,Carmi:2017jqz}.
See the generalizations of the conjectures to $CV 2.0$ and $CA
2.0$ in \cite{Tao:2017fsy,An:2018dbz,Fan:2018wnv}. Others see
\cite{Guo:2017rul,Ghaffarnejad:2018prc,Ghaffarnejad:2018bsd,Ghodrati:2017roz,Barbon:2018mxk,Hashimoto:2018bmb,Sinamuli:2018jhm,Karar:2018hvy,Susskind:2018fmx,Fu:2018kcp,Reynolds:2017jfs,Karar:2017org,Cottrell:2017ayj,Kim:2017qrq,Qaemmaqami:2017lzs,Nagasaki:2017kqe,Gan:2017qkz,Kim:2017lrw,Chapman:2016hwi}.

On the boundary field theory side, basically there are two ways to
understand the complexity of quantum fields. The one is
``path-integral complexity''
\cite{Czech:2017ryf,Caputa:2017urj,Caputa:2017yrh,Molina-Vilaplana:2018sfn,Bhattacharyya:2018wym}
and the other is ``geometric
complexity''\cite{Bhattacharyya:2018bbv,Caputa:2018kdj,Jefferson:2017sdb,Chapman:2017rqy,Yang:2017nfn,Khan:2018rzm,Yang:2018nda,Hackl:2018ptj,Alves:2018qfv,Magan:2018nmu},
based on different understandings on quantum gates in field
theory. Currently, one puzzle is that in most holographic work one
usually focuses on the evolution of the complexity beginning at a
TFD state, however in QFT one usually considers the evolution
relative to a vacuum state. These two processes may be different
in principle and further investigation is needed. Recently, the
complexity between the vacuum and the thermal state has been
studied by a holographic quench in Vaidya-AdS
spacetime\cite{Moosa:2017yvt,Ageev:2018nye,Chapman:2018dem,Chapman:2018lsv}.

Above CV and CA conjectures on the complexity are originally
proposed for global spacetime. Sequently they have been
generalized to be applicable for the subregion in
\cite{Alishahiha:2015rta} and \cite{Carmi:2016wjl}. Given a
boundary subregion $\mathcal{A}$ on a time slice
$\mathcal{\sigma}$, one can construct the corresponding
entanglement wedge $W[\mathcal{A}]$ and the
Wheeler-DeWitt patch $W_{WDW}[\mathcal{\sigma}]$. Then the
subregion CA conjecture tells us that the complexity of a boundary
state (which corresponds to the subregion $\mathcal{A}$) equals
the action of the intersecting region
$W[\mathcal{A}]\cap W_{WDW}[\mathcal{\sigma}]$. While the subregion CV conjecture tells us that the complexity of a boundary state is equal to the
volume of codimension-one extremal hypersurface
$\Gamma_{\mathcal{A}}$, which is bounded by the boundary
 subregion $\mathcal{A}$ and the corresponding
Hubeny-Rangamani-Takayanagi (HRT) surface $\gamma_{\mathcal{A}}$.
The formula is given by

\begin{equation}
  C_{\mathcal{A}}=\frac{V(\Gamma_{\mathcal{A}})}{8\pi l_{AdS}G_{N}}
\end{equation}
where $l_{AdS}$ is the AdS radius. In addition, some attempts
to understand the dual complexity of mixed states are recently
suggested in \cite{Agon:2018zso}. (See
\cite{Banerjee:2017qti,Zangeneh:2017tub,Bhattacharya:2018oeq,Zhang:2018qnt,Ben-Ami:2016qex,Roy:2017kha,Gan:2017qkz,Roy:2017uar,Abt:2017pmf,Chen:2018mcc,Du:2018uua,Abt:2018ywl}
for related works on the subregion complexity.)

The evolution of the holographic subregion complexity has been
investigated over the Vaidya-AdS spacetime in \cite{Chen:2018mcc}.
This dynamical process is dual to the
thermal quench in CFT on the boundary, and can be modelled
holographically by collapsing a thin shell of null matter from the
AdS boundary to form an AdS black brane. We intend to know more
details about this process and provide more physical understanding
on the results obtained in numerics. It is also desirable to
provide more information about the subregion complexity in the
boundary field theory.

In this paper we will explore the evolution of the subregion
complexity with CV conjecture over the background in
Einstein-Born-Infeld theory. The subregion we choose here is an
infinitely long strip with the width $l$. This evolution process
is dual to the process of a quench which is not only thermal, but
also electromagnetic in the sense that it is modelled
holographically by collapsing a null-like thin shell with mass $M$
and charge $Q$ from the AdS boundary to form Born-Infeld-AdS
(BI-AdS) black brane \cite{Camilo:2014npa}.

Born-Infeld (BI) electrodynamics was firstly introduced by Born and Infeld in the 1930's \cite{Born:1934gh}. They proposed a non-linear modification to Maxwell theory to regularize the divergence of self-energy of a point-like charged particle. Recently BI electrodynamics becomes more intriguing in superstring theory. The low energy behavior of the vector modes of open strings and dynamics of D-branes are given by the BI action and its similar non-Abelian version respectively.(See \cite{Gibbons:2001gy,deAlwis:1998mi} for related works.) Further, BI theory plays an important role in the modified gravity \cite{BeltranJimenez:2017doy} and inspires a new approach to avoid spacetime spacetime singularities in the high energy or highly curved regime.
 
Here we desire to capture more general features of the evolution of complexity caused by BI electrodynamics that may not appear in Einstein theory. On the one hand, in the limit of $b\rightarrow 0$, this framework covers Einstein-Maxwell theory and can be applied to explore the thermalization process of a strongly coupled system with chemical potential. In \cite{Caceres:2012em}, it was found that the thermalization time for the two-point function, Wilson loop and entanglement entropy are increased with the charge (or the chemical potential). Thus it also deserves to find out the effect of chemical potential on the subregion complexity. On the other hand, for non-zero parameter $b$ more interesting effects caused by the nonlinearity of electrodynamics will be disclosed. The parameter $b$ was found to have the opposite effect against the charge $Q$. That is to say, as the parameter $b$ grows, the thermalization time for above non-local probes is decreased (see \cite{Camilo:2014npa}). Inspired by the former works, we intend to explore what will happen to the subregion complexity during the holographic quench process with nonlinear electrodynamics.

We organize the paper as follows. In Section 2, we introduce
the general setup for the bulk geometry with Vaidya-type black brane solutions in Einstein-Born-Infeld theory. Then we
derive the holographic entanglement entropy(HEE) and the subregion
complexity for a strip on the boundary. In Section 3 we
numerically calculate the evolution of the holographic
subregion complexity as well as the holographic entanglement
entropy. The impact of the charge $Q$ and the inverse of BI
parameter $b$ on the evolution is investigated. Section 4 is the
conclusions and outlooks.

\section{The Setup}

In this section we will briefly review the Einstein-Born-Infeld
theory which contains a non-linear term of electrodynamics, and
then present a Vaidya-type black brane background, which is
holographically dual to the quench process from a vacuum state to
a thermal state on the boundary. Given a strip on the boundary,
we will derive the analytical expressions for its HEE and the
holographic subregion complexity.

\subsection{Einstein-Born-Infeld Theory}

The action for $(d+1)$-dimensional Einstein gravity minimally
coupled to Born-Infeld electrodynamics can be expressed
as\cite{Dey:2004yt} (see also \cite{Cai:2004eh,Li:2016nll,Cai:2017sjv})
\begin{equation}\label{eq:EBIaction}
  S = \frac{1}{16\pi G}\int d^{d+1}x \sqrt{-g}\big[R-2\Lambda+L_{BI}(F)\big]\ ,
 \end{equation}
 where $L_{BI}(F)$ is given by
 \begin{equation}
  L_{BI}(F) = 4b^{-2}\left(1-\sqrt{1+\frac{F_{\mu\nu}F^{\mu\nu}}{2}b^{2}}\right)\ .
 \end{equation}
The constant $b$ is the inverse of the ordinary BI parameter
(for numerical convenience). In the limit $b\rightarrow 0$, the
action goes back to Einstein-Maxwell theory. And here
we choose $16\pi G=1$. The metric of BI-AdS solution with a
planar horizon can be expressed as
\begin{equation}\label{eq:metricBB}
  ds^2 = -U(r)dt^2+\frac{dr^2}{U(r)}+ r^2\sum_{i=1}^{d-1}dx_i^2\ ,
\end{equation}
where
\begin{align}\label{eq:V(r)}
  U(r)=& -\frac{M}{r^{d-2}}+\left[\frac{4b^{-2}}{d(d-1)}+1\right]r^2-\frac{2\sqrt{2}b^{-1}}{d(d-1)r^{d-3}}\sqrt{2b^{-2}r^{2d-2}+(d-1)(d-2)Q^2}\nonumber\\
  &+\frac{2(d-1)Q^2}{dr^{2d-4}}{}_{2}F_1\left[\frac{d-2}{2d-2},\frac{1}{2};\frac{3d-4}{2d-2};-\frac{(d-1)(d-2)Q^2b^{2}}{2r^{2d-2}}\right],
\end{align}
and the AdS radius is set to $1$. The event horizon is defined by
$U(r_h)=0$ and since the horizon is planar, we should regard this
spacetime as a black brane as mentioned in \cite{Camilo:2014npa}.
In the next subsection we will generalize it to a
time-dependent background which is so-called the Vaiyda-BI-AdS
spacetime.

\subsubsection{Vaidya-BI-AdS Metric}

To obtain Vaidya-BI-AdS metric, we firstly rewrite the metric
(\ref{eq:metricBB}) in Eddington-Finkelstein coordinate system by
the following transformations
\begin{align}
  dv=dt+dr/U(r),\nonumber\\
  z=1/r.\nonumber
\end{align}
Then the metric is expressed as
\begin{equation}\label{eq:metricBBEF}
  ds^2 = \frac{1}{z^2}\left[-f(z)dv^2-2dvdz+\sum_{i=1}^{d-1}dx_i^2\right]\ ,
\end{equation}
where
\begin{equation}
  f(z) = z^2 U\left(\frac{1}{z}\right) .
 \end{equation}
In addition, from the metric in (\ref{eq:metricBB}), one can
derive the Hawking temperature as
\begin{equation}\label{eq:THawking}
  T = \frac{1}{4\pi r_h}\left[\left(\frac{4b^{-2}}{d-1}+d\right)r_h^2-\frac{2\sqrt{2}b^{-1}}{(d-1)r_h^{d-3}}\sqrt{2b^{-2}r_h^{2d-2}+(d-1)(d-2)Q^2}\right] .
 \end{equation}
In particular, when Hawking temperature $T=0$, we obtain an
extremal black brane. Under this condition, the charge takes the
maximal value $Q=Q_{ext}$ which is
\begin{equation}\label{eq:Qext}
  Q_{ext}^2 = \frac{d}{(d-2)}\left[1+\frac{d(d-1)b^2}{8}\right]r_h^{2d-2} .
 \end{equation}
Now we extend it to the Vaidya-BI-AdS metric in which both the
mass and the charge of the black brane are treated as functions
of $v$. That is
 \begin{align}
  m(v)=\frac{M}{2}\left(1+\tanh\frac{v}{v_{0}}\right)\label{eq:Massfunction}, \\
  q(v)=\frac{Q}{2}\left(1+\tanh\frac{v}{v_{0}}\right)\nonumber,
\end{align}
where $M$ and $Q$ are the parameters of the BI-AdS black brane and
$v_0$ denotes the thickness of the shell. This extension leads
to the following dynamical background
\begin{equation}
  ds^{2}=\frac{1}{z^{2}}\left[-f(v,z)dv^{2}-2dzdv+dx^{2}+\sum_{i=1}^{d-2}dy_{i}^{2}\right],\label{eq:VaidyaMetric}
  \end{equation}
  \begin{align}
    f(v,z)=&1+\frac{2(d-1)}{d}{}_2 F_1\left[\frac{1}{2},\frac{d-2}{2d-2},\frac{4-3d}{2-2d},-\frac{b^2}{2}(d-2)(d-1)q(v)^2 z^{2(d-1)}\right]q(v)^2 z^{2(d-1)} \nonumber\\
    &+\frac{4}{b^2d(d-1)}-m(v)z^d-\frac{2z^{d-1}}{b(d-1)d}\sqrt{2(2-3d+d^2)q(v)^2+\frac{4z^{2-2d}}{b^2}}. \nonumber
  \end{align}

Since in Eq.(\ref{eq:Massfunction}) we have changed mass
$M$ and charge $Q$ into a time-dependent form, it is obvious
that the metric in Eq.(\ref{eq:VaidyaMetric}) is not a solution of
the original action as shown in Eq.(\ref{eq:EBIaction}).
Therefore, to guarantee that Eq.(\ref{eq:Massfunction}) could
be a solution to Einstein equations, we need add some external
source term $S_{ex}$ to provide a variation of $M$ and $Q$. Taking
$S_{ex}$ into account, the equations of motion can be expressed as
  \begin{align}
    R_{\mu\nu}-\frac{1}{2}Rg_{\mu\nu}-2b^{-2}g_{\mu\nu}\left(1-\sqrt{1+b^2F^2/2}\right)-\frac{2F_{\mu\rho}F_{\nu}{}^{\rho}}{\sqrt{1+b^2F^2/2}}&=-8\pi GT_{\mu\nu}^{(ex)},\\
\nabla_{\mu}\left(\frac{F^{\mu\nu}}{\sqrt{1+b^2F^2/2}}\right)&=-8\pi GJ^{\nu}_{(ex)}.
  \end{align}
Here we keep the cosmological constant $\Lambda$ and the
Newton's constant $G$ temporarily in the above equations. Then,
the corresponding source $T_{\mu\nu}$ and $J^{\nu}$ can be solved
as
  \begin{align}
    T_{\mu\nu}^{(ex)}=&\frac{d-1}{z^{1-d}}\left[\dot{m}(v)-\frac{2}{z^{2-d}}{}_2F_1\left[\frac{d-2}{2d-2};\frac{1}{2};\frac{3d-4}{2d-2};\frac{(d-2)(d-2)q(v)^2}{-2b^{-2}z^{2-2d}}\right]q(v)\dot{q}(v)\right]\delta_\mu^v\delta_\nu^v,\\
    J^{\nu}_{(ex)}=&\sqrt{\frac{(d-1)(d-2)}{8}}z^{d+1}\dot{q}(v)\delta^{\nu v},
  \end{align}
where we denote the dot as $\partial_v$ and reset $16\pi G=1$
as well as the AdS radius.
\subsection{Holographic description of entanglement and complexity for a
strip}

In this subsection we analytically derive the integral
expressions of holographic entanglement entropy and complexity for
a $(d-1)$-dimensional strip $\mathcal{A}$ on the boundary. The
strip can be parameterized by the boundary coordinates
$(x,y_{1},...,y_{d-2})$. We further assume that it has a width of
$l$ along $x$ direction such that $x\in [-l/2,l/2]$, while it has
infinite length along the directions of $y_i$ such that $y_{i}\in
(-\infty,\infty)$, where $i=1,...,d-2$. We will figure out the HRT
surface $\gamma_{\mathcal{A}}$ at first, and then locate the
codimension-one extremal surface $\Gamma_{\mathcal{A}}$ such that
the evolution of the holographic subregion complexity can be
evaluated by subregion CV conjecture.

\subsubsection{Holographic Entanglement Entropy}

Given a strip, the corresponding HRT surface can be parameterized
by $z(x)$ and $v(x)$,with the boundary conditions
\begin{equation}
  z(-l/2)=z(l/2)=\epsilon, v(-l/2)=v(l/2)=t-\epsilon,
\end{equation}
where $\epsilon$ is a cut-off constant. At the tip of the HRT
surface we have
\begin{equation}
  z'(0)=v'(0)=0, z(0)=z_{t}, v(0)=v_{t},
\end{equation}
where $(z_{t}$, $v_{t})$ label the location of the tip and
also characterize the HRT surface at boundary time $t$. As shown
in \cite{Chen:2018mcc}, the induced metric on the HRT surface
has the form as
\begin{equation}
  ds^{2}=\frac{1}{z^{2}}\left[-f(v,z)v'^{2}-2z'v'+1\right]dx^{2}+\frac{1}{z^{2}}\sum_{i=1}^{d-2}dy_{i}^{2}.
\end{equation}
The area of the HRT surface $\gamma_{\mathcal{A}}$ is
\begin{equation}\label{eq:Vga}
  A_{t}(\gamma_{\mathcal{A}})=L^{d-2}\int_{-l/2}^{l/2}\frac{\sqrt{1-f(v,z)v'^{2}-2z'v'}}{z^{d-1}}dx,
  \end{equation}
where $t$ denotes the HRT surface which is anchored on a boundary
time slice with time $t$ and $L^{d-2}$ is the infinite area
related to directions $y_{i}$. Treating the area functional
$A_{t}(\gamma_{\mathcal{A}})$ as an action we can read the
Lagrangian and the corresponding Hamiltonian as
\begin{align}
  \mathcal{L}_{S}=\frac{\sqrt{1-f(v,z)v'^{2}-2z'v'}}{z^{d-1}},\\
  \mathcal{H}_{S}=\frac{1}{z^{d-1}\sqrt{1-f(v,z)v'^{2}-2z'v'}}.
\end{align}
Since the Hamiltonian is conserved along the direction $x$, we have
\begin{equation}
  1-f(v,z)v'^{2}-2z'v'=\frac{z_{t}^{2d-2}}{z^{2d-2}}.\label{eq:RTt}
\end{equation}
Then we take the derivative of (\ref{eq:RTt}) and substitute it into
the equations of motion (E.O.M) of $z(x)$ and $v(x)$
respectively, leading to
\begin{align}
  0= & -2(d-1)+2zv''+v'\left[2(d-1)f(v,z)v'+4(d-1)z'-zv'\partial_{z}f(v,z)\right],\label{eq:vpp}\\
  0= & 2(d-1)f(v,z)^{2}v'^{2}+f(v,z)\left[-2(d-1)+4(d-1)v'z'-zv'^{2}\partial_{z}f(v,z)\right]\label{eq:zpp}\\
 & -z\left[2z''+v'\left(2z'\partial_{z}f(v,z)+v'\partial_{v}f(v,z)\right)\right].\nonumber
\end{align}

We numerically solve above equations for the HRT surface
$\gamma_{\mathcal{A}}$ and denote the solutions as
$(\tilde{v}(x),\tilde{z}(x))$, then the equation in
(\ref{eq:Vga}) becomes
\begin{equation}
  A_{t}(\gamma_{\mathcal{A}})=2L^{d-2}\int_{0}^{l/2}\frac{z_{t}^{d-1}}{\tilde{z}(x)^{2d-2}}dx.
\end{equation}
It corresponds to the holographic entanglement entropy of the
strip on the boundary. Next we need to work out the solution of
the codimension-one extremal surface $\Gamma_{\mathcal{A}}$ at
various boundary time $t$ to study the evolution behavior of
the holographic subregion complexity.

\subsubsection{Holographic Subregion Complexity}

Recall that the codimension-one extremal surface
$\Gamma_{\mathcal{A}}$ is bounded by $\mathcal{A}$ on the boundary
and the HRT surface $\gamma_{\mathcal{A}}$ in the bulk. As
suggested in \cite{Chen:2018mcc}, $\Gamma_{\mathcal{A}}$ can be
parameterized by $z(v,x)$ in general. For this model thanks to
the translational invariance, the extremal surface
$\Gamma_{\mathcal{A}}$ is independent of $x$, so the
parameterization can simply be written as

\begin{equation}
  z=z(v).
\end{equation}
As a result, the induced metric on the extremal surface
$\Gamma_{\mathcal{A}}$ is
\begin{equation}
  ds^{2}=\frac{1}{z^{2}}\left[-\left(f(v,z)+2\frac{\partial z}{\partial
  v}\right)dv^{2}+dx^{2}+\sum_{i=1}^{d-2}dy_{i}^{2}\right],
\end{equation}
and the volume of $\Gamma_{\mathcal{A}}$ is given by
\begin{equation}\label{eq:VGa}
  V_{t}(\Gamma_{\mathcal{A}})=2L^{d-2}\int_{v_{t}}^{\tilde{v}(l/2)}dv\int_{0}^{\tilde{x}(v)}dx\left[-f(v,z)-2\frac{\partial z}{\partial v}\right]^{1/2}z^{-d},
\end{equation}
where $\tilde{x}(v)$ is the $x$ coordinate on the HRT surface
$\gamma_{\mathcal{A}}$. Then we can write down the Lagrangian
\begin{equation}
  \mathcal{L}_{V}=\left[-f(v,z)-2\frac{\partial z}{\partial
  v}\right]^{1/2}z^{-d},
\end{equation}
and the corresponding E.O.M of\footnote{In \cite{Chen:2018mcc},
the equation contains typing errors.} $z(v)$
\begin{align}
  0=&[2df(v,z)^{2}+4dz'(v)^{2}-3z(v)z'(v)\partial_{z}f(v,z)+f(v,z)(6dz'(v)-z(v)\partial_{z}f(v,z))\nonumber\\
  &-z(v)(2z''(v)+\partial_{v}f(v,z))]/[z(v)^{1+d}(-f(v,z)-2z'(v))^{3/2}].\label{eq:VolumeEqZV}
\end{align}
In principle, one should solve E.O.M (\ref{eq:VolumeEqZV}) for
$z(v)$, with boundary conditions determined by
$\gamma_{\mathcal{A}}$ and $\mathcal{A}$. However as proved in
\cite{Chen:2018mcc}, the relation $\tilde{z}(\tilde{v})$ (where
$\tilde{z}$ and $\tilde{v}$ are the solutions for the HRT
surface $\gamma_{\mathcal{A}}$) is just the solution of the E.O.M
(\ref{eq:VolumeEqZV}). Thus the equation in (\ref{eq:VGa})
becomes
\begin{equation}
  V_{t}(\Gamma_{\mathcal{A}})=2L^{d-2}\int_{v_{t}}^{\tilde{v}(l/2)}dv\left[-f(v,z(v))-2\frac{\partial z}{\partial v}\right]^{1/2}z(v)^{-d}\tilde{x}(v).\label{eq:VolumeZV}
\end{equation}

So far, for a given strip on the boundary, we have figured out the
integral expressions of the HRT surface $\gamma_{\mathcal{A}}$ and
the codimension-one extremal surface $\Gamma_{\mathcal{A}}$ at
some boundary time $t$. In next section we will explore the
evolution behavior of holographic entanglement entropy and the
subregion complexity in numerical manner.

\section{Holographic Subregion Complexity in Einstein-Born-Infeld Theory}

The quench in CFT could be described holographically by the
evolution of the bulk geometry in Einstein-Born-Infeld theory,
whose initial state corresponds to the pure AdS and final state
corresponds to the BI-AdS black brane. In this section we
first work out the evolution of the holographic entanglement
entropy, and then explore the evolution of the subregion
complexity numerically after the global quench. Afterwards, we
study the effect of the charge $Q$ and the parameter $b$ on
the evolution of the subregion complexity.

\subsection{Numeric Setup}

\begin{figure}
  \centering
\subfigure[]{\label{l=1zt}
\includegraphics[width=145pt]{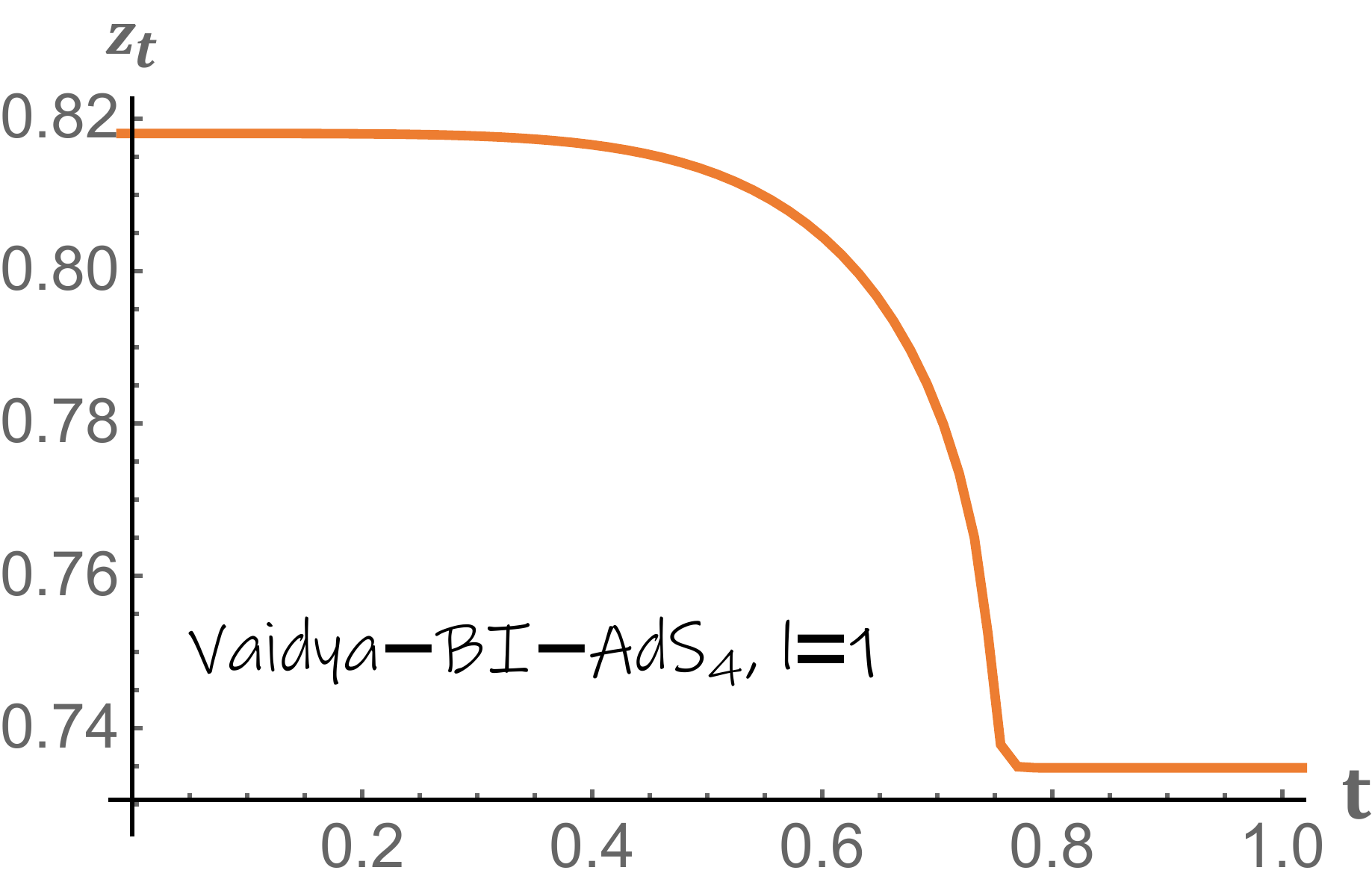}}
\hspace{0pt}
\subfigure[]{\label{l=1s}
\includegraphics[width=145pt]{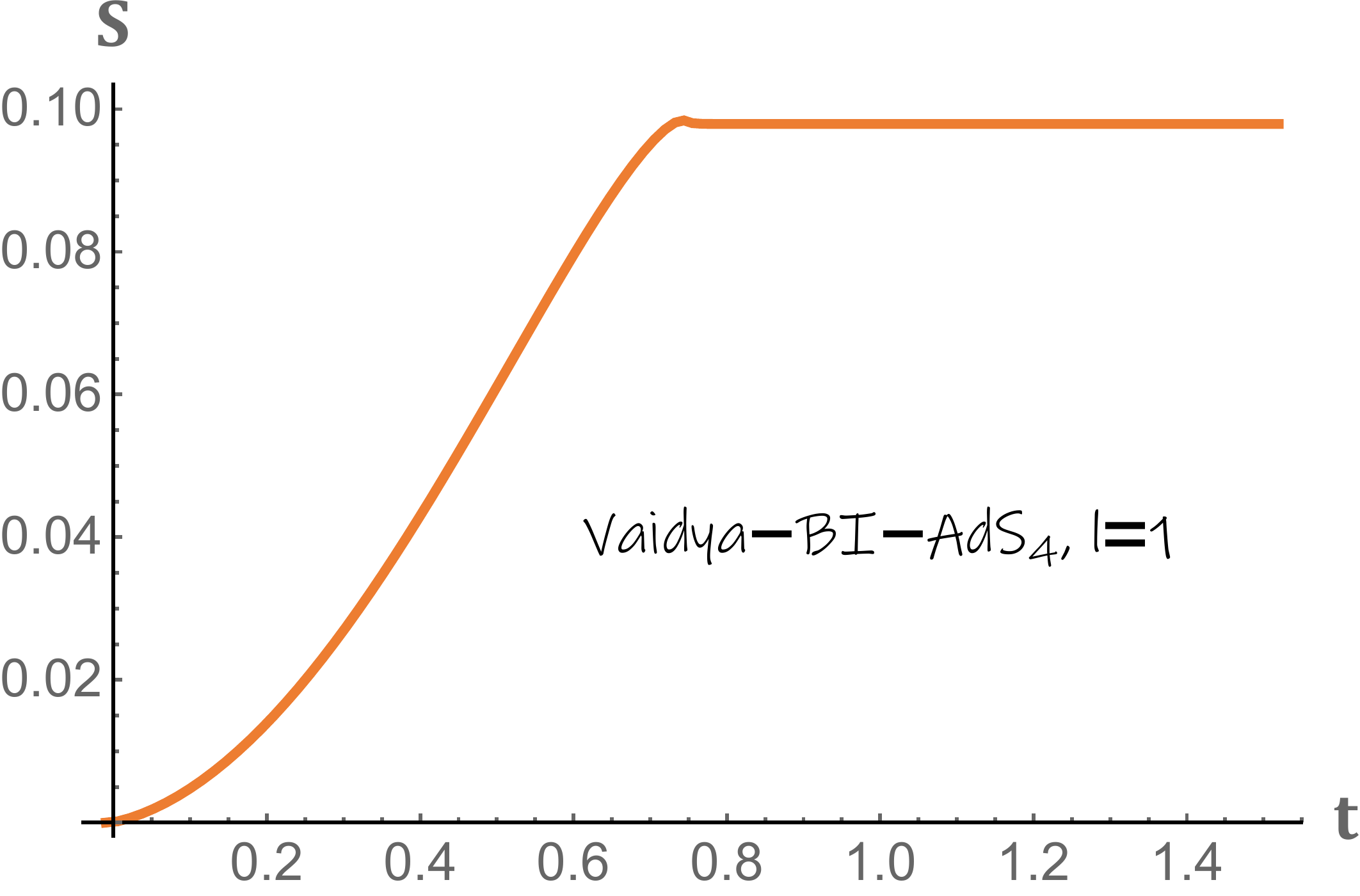}}
\hspace{0pt}
\subfigure[]{\label{l=1c}
\includegraphics[width=145pt]{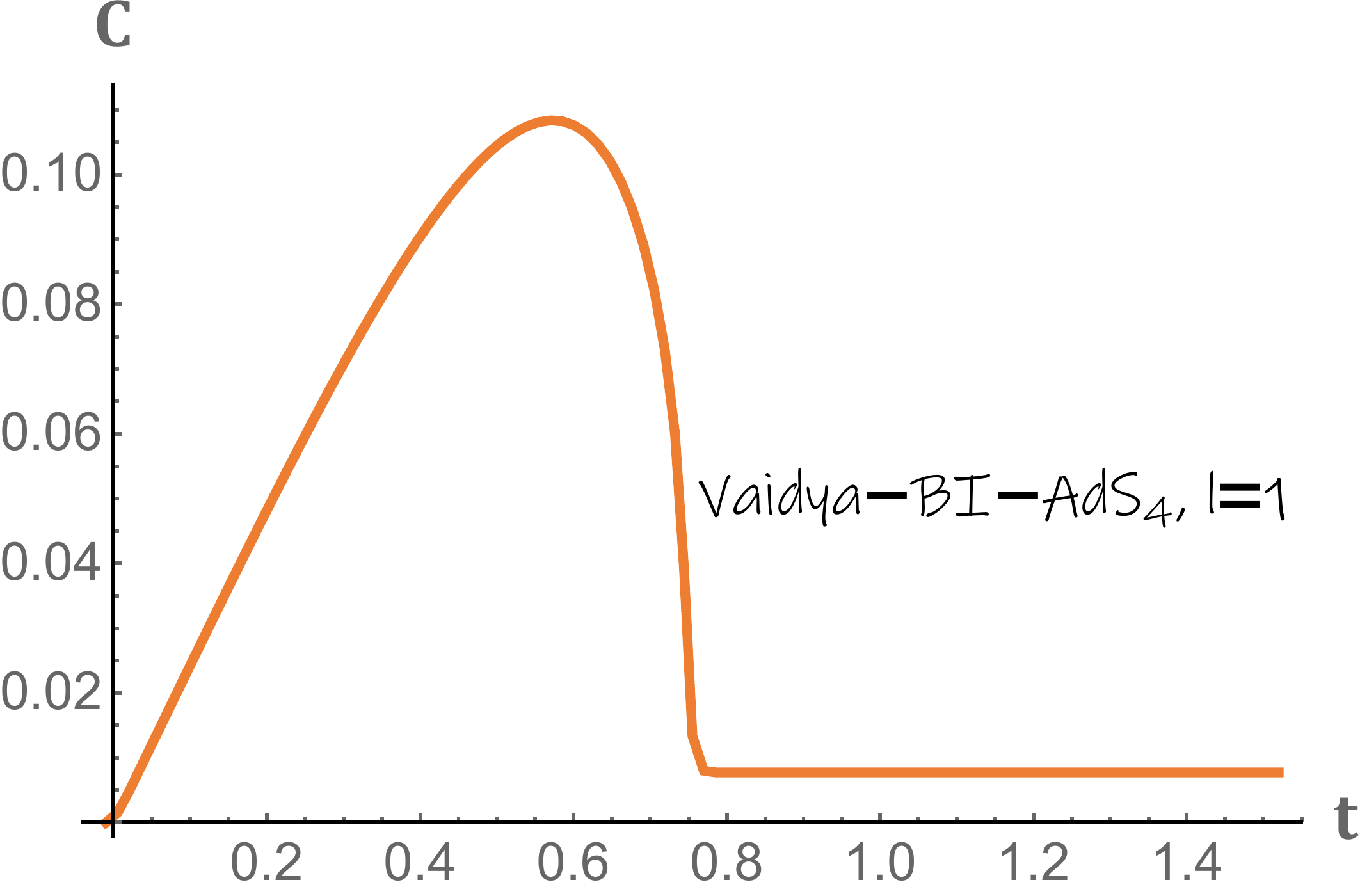}}
\hspace{0pt}
\subfigure[]{\label{l=5zt}
\includegraphics[width=145pt]{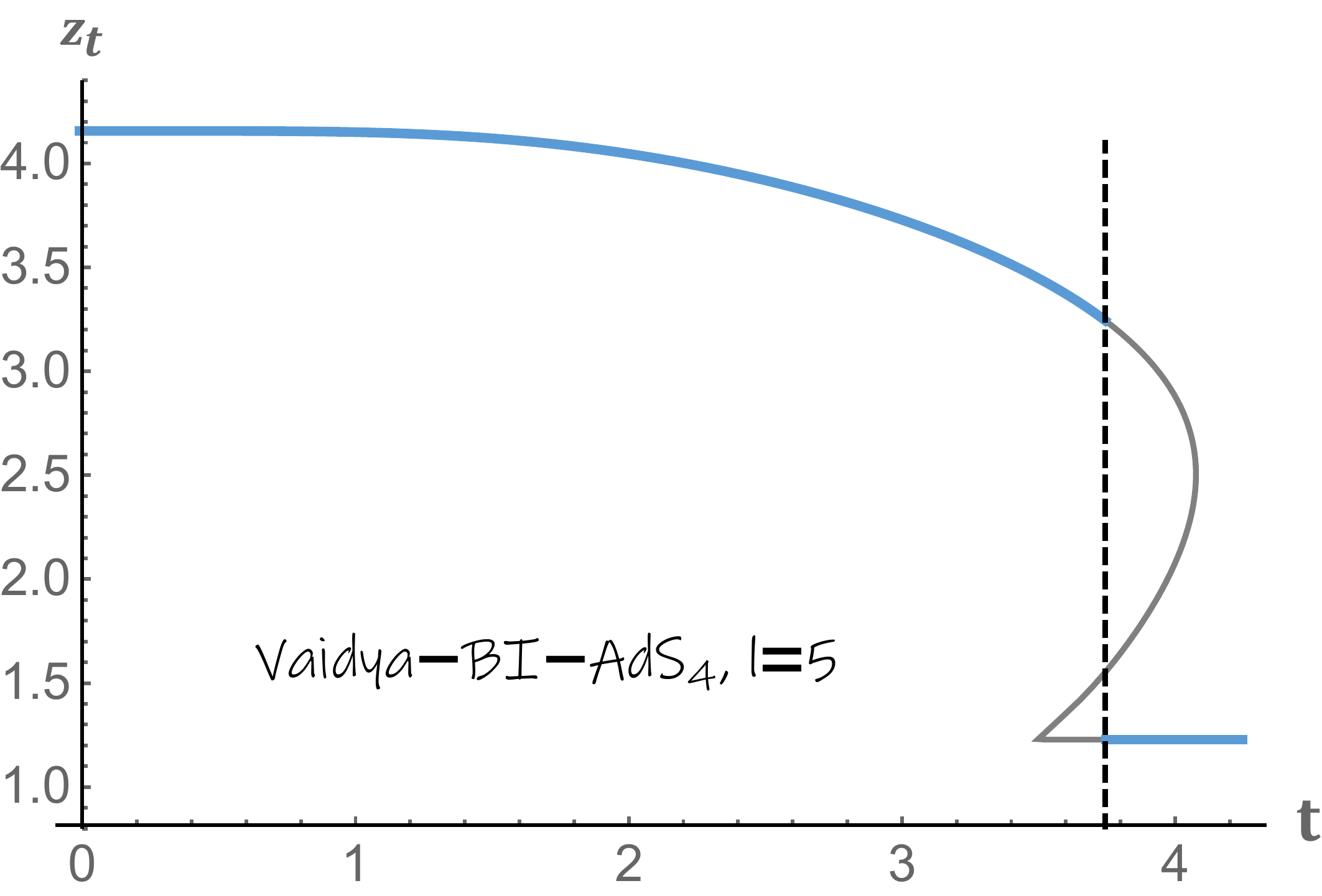}}
\hspace{0pt}
\subfigure[]{\label{l=5s}
\includegraphics[width=145pt]{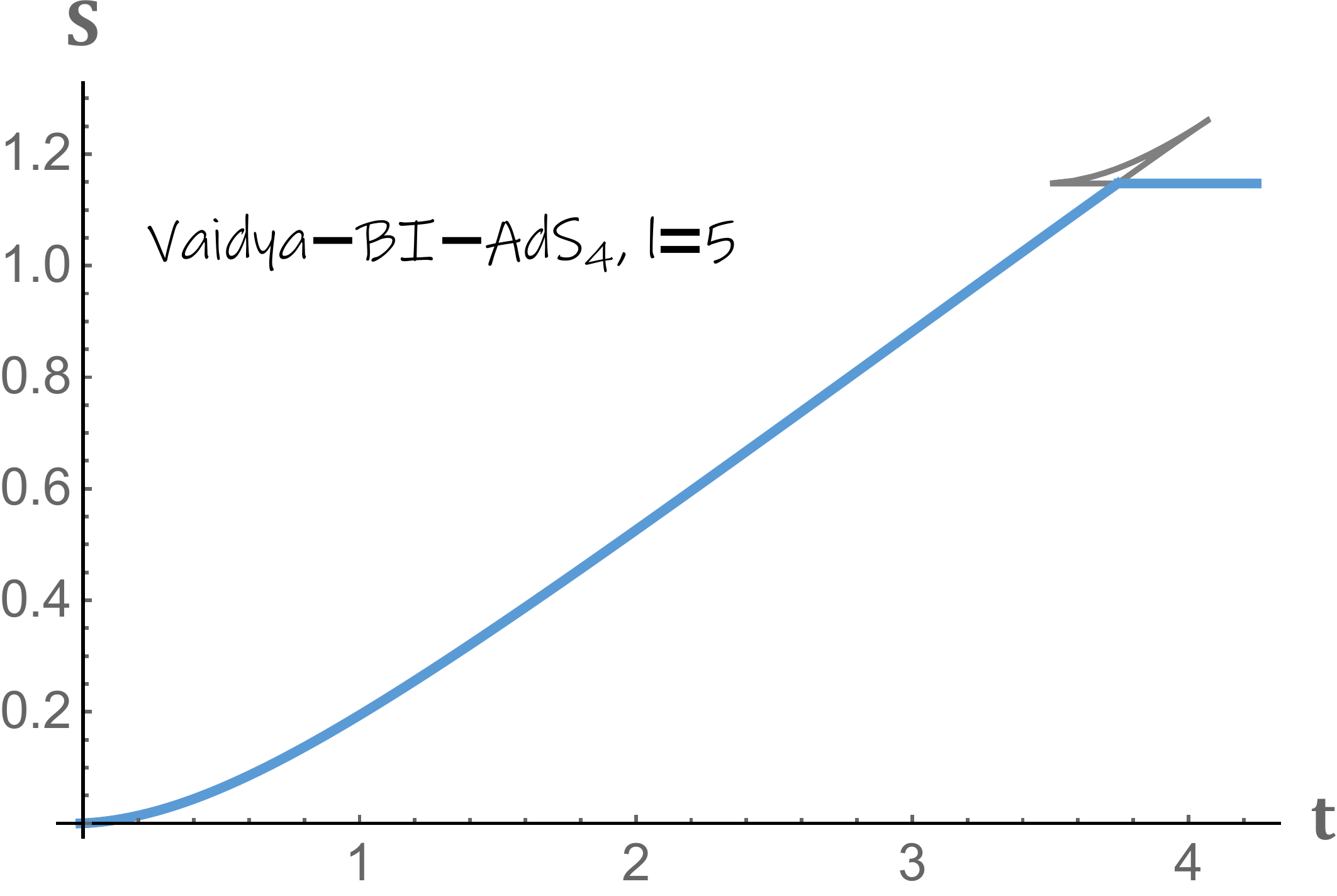}}
\hspace{0pt}
\subfigure[]{\label{l=5c}
\includegraphics[width=145pt]{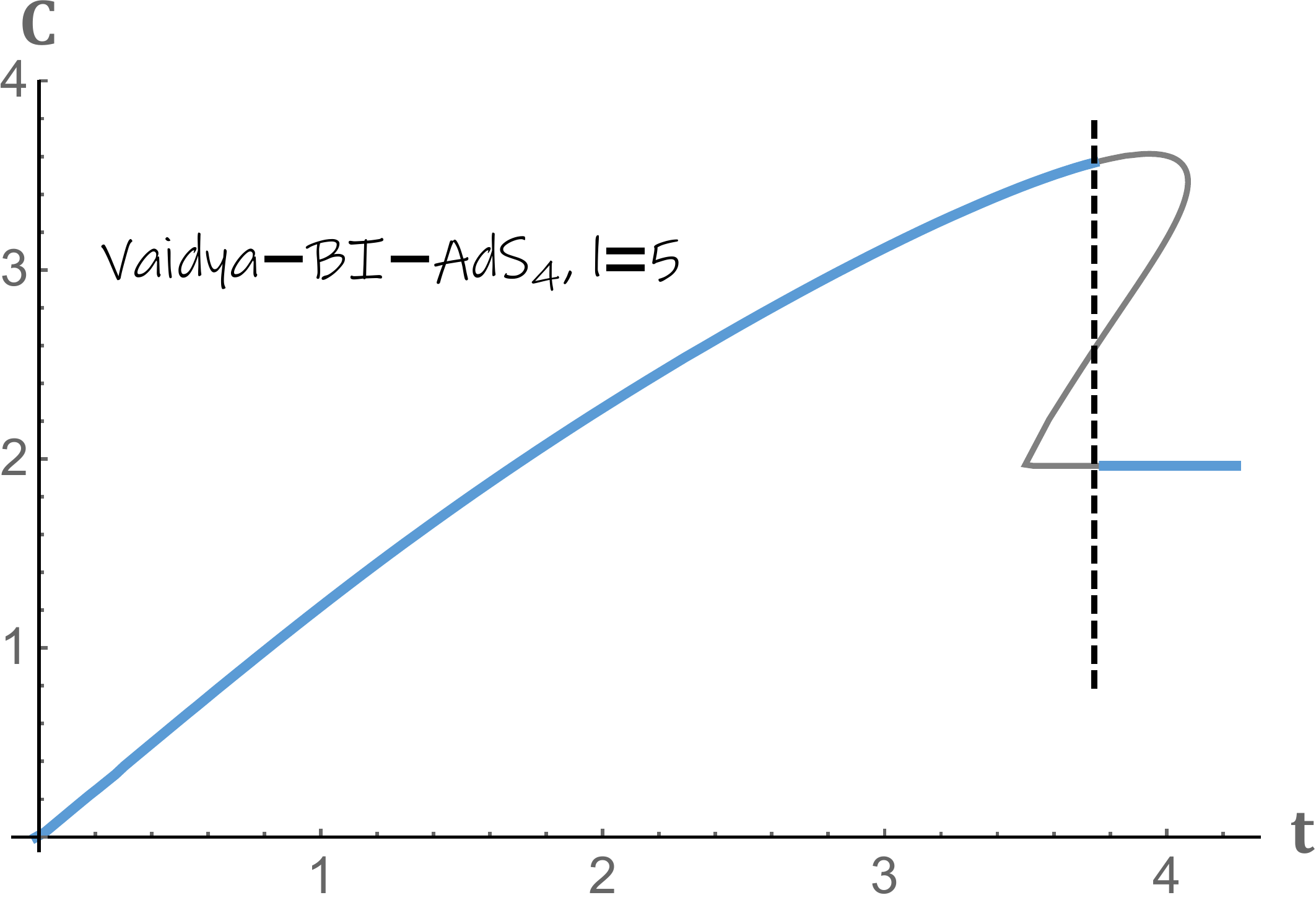}}

\caption{Two distinct patterns of the evolution of HEE and
subregion complexity. The figures on the top display a continuous
pattern with $l=1$, while the figures at the bottom display a
discontinuous pattern with $l=5$. The dashed lines denote the
critical time for the transition $t_{c}=3.7434$.
 }\label{l=1&l=5}
\end{figure}

For numerical analysis we need to fix all free parameters and get
rid of the UV divergence. Here we take the UV cut-off
$z_{\epsilon}$ to be $\frac{1}{20}$, which turns out to be good
enough for us to obtain the cut-off independent data. The
thickness of the shell $v_0$ is set to be $\frac{1}{100}$, the
boundary dimension $d$ to be $3$ if without notice(which means
we mainly focus on $AdS_{4}/CFT_{3}$) and the mass $M$ of the
final black brane to be $1$. With this setup equations
(\ref{eq:THawking}) and (\ref{eq:Qext}) reduce to

\begin{equation}\label{eq:THawking1}
  T = \frac{1}{4\pi r_h}\left[(2b^{-2}+3)r_h^2-2b^{-1}\sqrt{b^{-2}r_h^{4}+Q^2}\right]
 \end{equation}
and
\begin{equation}\label{eq:Qext1}
  Q_{ext}= \sqrt{3\left(1+\frac{3b^2}{4}\right)r_h^{4}}.
 \end{equation}
Given the value of parameter $b$, we can set the charge $Q\in
[0,Q_{ext}]$ to explore the evolution of the holographic
entanglement entropy as well as subregion complexity.

Next we will solve E.O.M (\ref{eq:zpp}) and (\ref{eq:vpp})
for $(\tilde{v}(x),\tilde{z}(x))$ with the boundary conditions
\begin{equation}
  v'(0)=z'(0)=0,\ \ z(0)=z_{t},\ v(0)=v_{t},
\end{equation}
by the shooting method.

\subsection{The Evolution of Subregion Complexity}

\begin{figure}
  \centering
  \subfigure[]{\label{l=5&Q=0.65&S}
  \includegraphics[width=200pt]{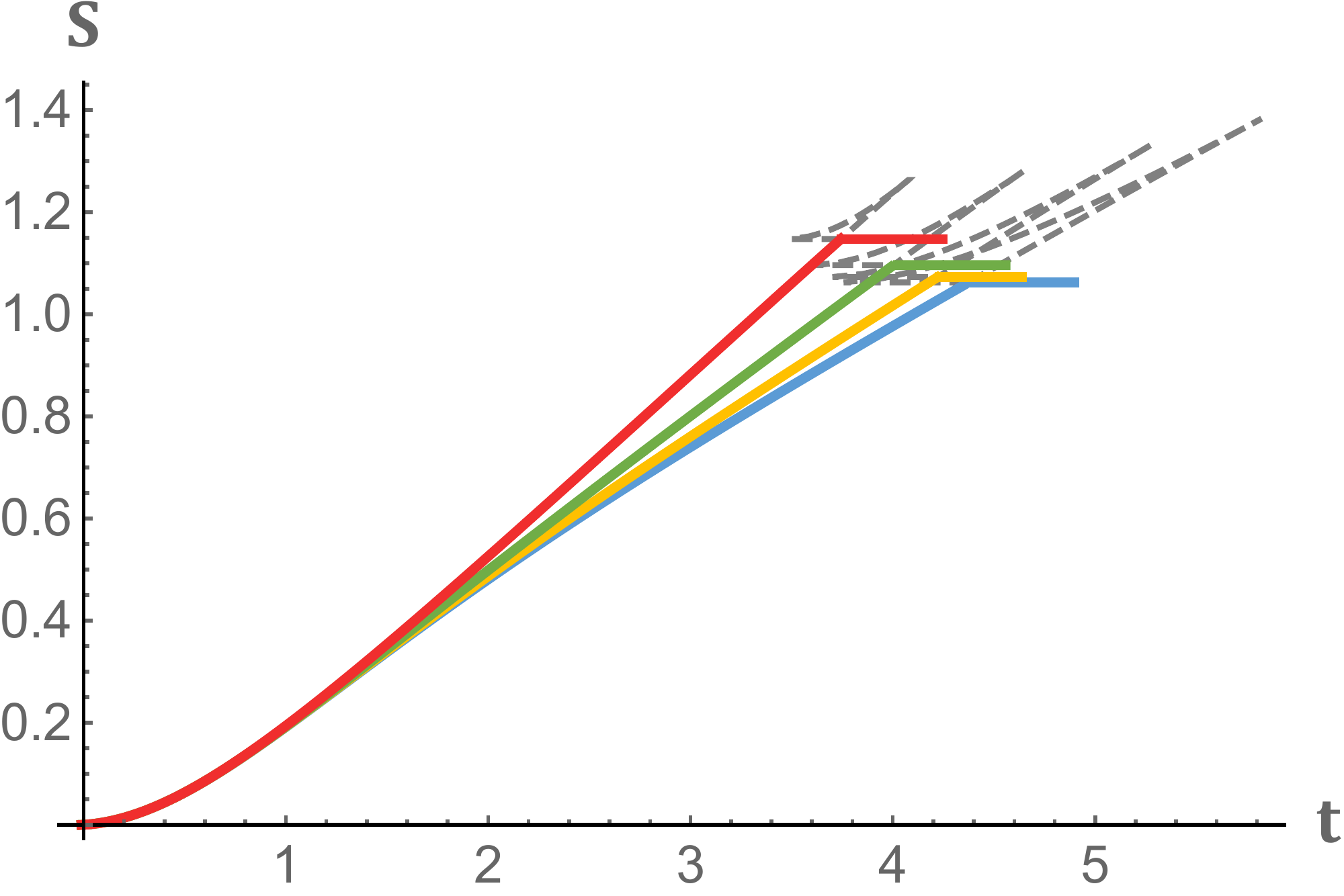}}
  \hspace{0pt}
  \subfigure[]{\label{l=5&b=2&S}
  \includegraphics[width=200pt]{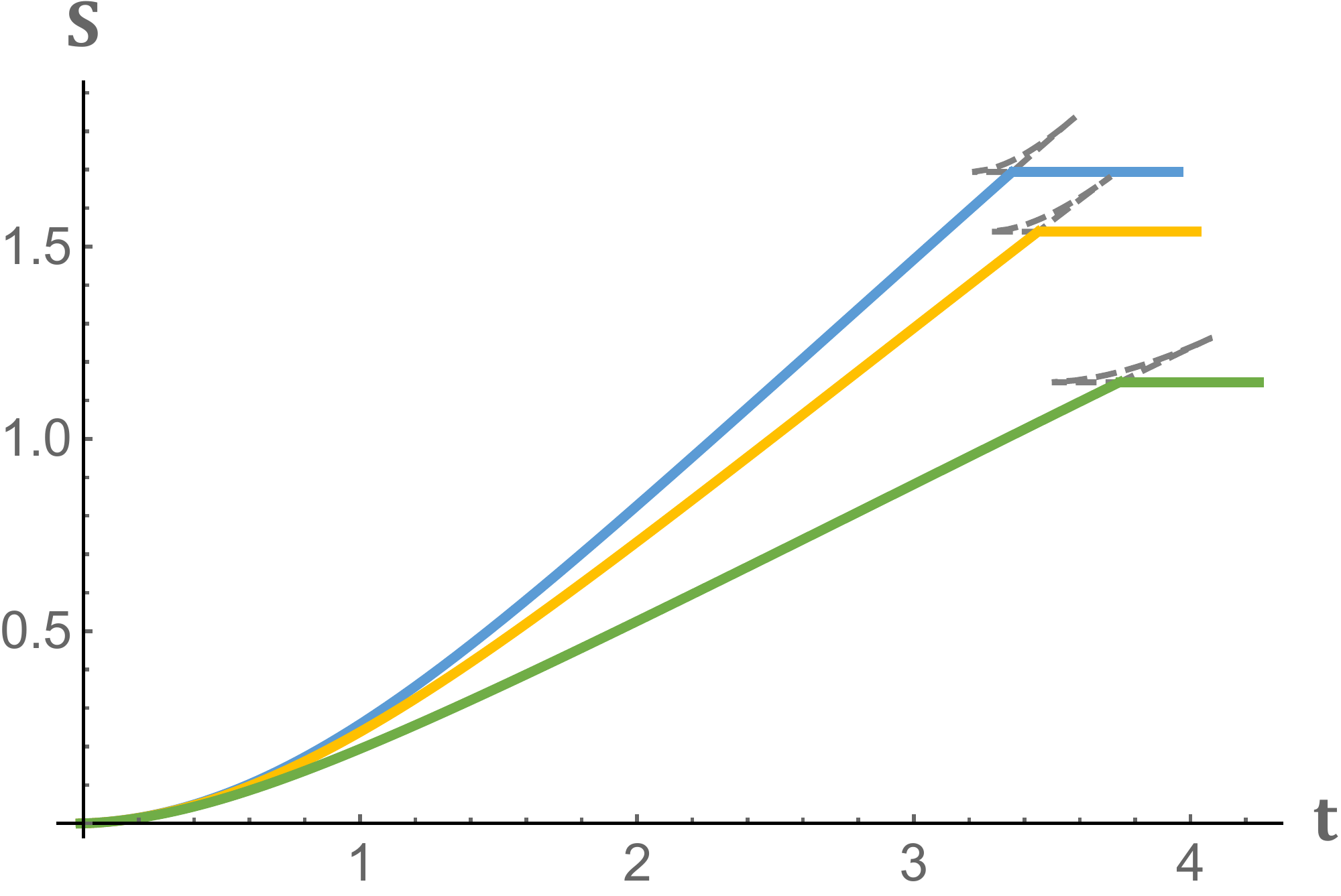}}
\caption{The left plot is for the evolution of HEE with various
values of $b$, where the blue, yellow, green and red lines
correspond to $b=0$, $\frac{1}{2}$, $1$, $2$, respectively. The
charge and the width of the strip are fixed as $Q=0.65$, $l=5$.
While the right plot is for the evolution of HEE with various
values of $Q$, where the blue, yellow and green lines correspond
to $Q=0.2$, $0.4$, $0.65$, respectively.
}\label{S}
\end{figure}

Once we figure out the HRT surface, the corresponding HEE can
be obtained from equation (\ref{eq:Vga}). Since we are only
concerned with the change of the HEE during the quench, we may
subtract the vacuum HEE and define a finite quantity for HEE as
\begin{equation}
  S=\frac{A_{t}(\gamma_{\mathcal{A}})-A_{AdS}(\gamma_{\mathcal{A}})}{2L^{d-2}}.
\end{equation}

Furthermore, the holographic subregion complexity can be
obtained by computing the volume of the codimension-one surface
$\Gamma_{\mathcal{A}}$ from equation (\ref{eq:VGa}). In parallel,
we define a normalized expression for the subregion complexity as
\begin{equation}
C=\frac{V_{t}(\Gamma_{\mathcal{A}})-V_{AdS}(\Gamma_{\mathcal{A}})}{2L^{d-2}}.
\end{equation}

Next we present our numerical results for the time evolution of
these two quantities during the course of the quench. In
Fig.\ref{l=1&l=5} we demonstrate two typical patterns of
evolution: the continuous pattern and the discontinuous pattern.
We choose the same charge $Q=0.65$ (which is less than the
extremal charge) and the same parameter $b=2$ but different width
$l=1$ and $l=5$ respectively. First of all, from Fig.\ref{l=1zt}
and Fig.\ref{l=5zt} we learn that the tip $z_t$ of the HRT surface
$\gamma_{\mathcal{A}}$ is decreasing with the time and finally
reaches a constant. In comparison, we notice that the HRT
surface with large $l$ takes longer time to get stable than
the HRT surface with smaller $l$. This phenomenon can be
intuitively understood based on the previous work in
\cite{Liu:2013qca} and \cite{Liu:2013iza}. During the entanglement
tsunami, the infalling thin shell divides the spacetime into two
parts, namely the AdS-Schwarzschild and the pure AdS. The former
region is swept by the tsunami while the latter region has not
been affected by the tsunami yet. As a result, the HRT surface is
also divided into two parts. The part in the AdS region is located
on a time slice just like the static case, while the other part in
the Schwarzschild region is not. Because the tip of the HRT
surfaces with large $l$ stretches deeper into the bulk, the
infalling shell need take longer time to reach this location.
Therefore, the location of HRT surfaces with larger $l$ will take
longer time to get stable.

Secondly, we notice that when $l$ is large ($l=5$), the HEE
evolution will display a swallow tail before getting stable,
which is marked by the gray line. This phenomenon has previously
been observed in \cite{Albash:2010mv} as
well. It indicates that at some given boundary time $t$, there
exist multi solutions for the surface $\gamma_{\mathcal{A}}$. We
only keep the solutions with minimum area as the HEE.

In addition, the growth rate of HEE depends on the charge $Q$
and the parameter $b$, as shown in Fig.\ref{S}.
Fig.\ref{l=5&Q=0.65&S} demonstrates the evolution of HEE with
different values of $b$. It is noticed that the larger the
parameter $b$ is, the sooner the curve saturates and the larger
the maximal value is. While Fig.\ref{l=5&b=2&S} shows the growth
curves with various values of $Q$. We find that the larger the
charge $Q$ is, the later the curve saturates and the smaller the
maximal value is.

Next we turn to the evolution of the subregion complexity. In
general, we observe that it increases in the early stage of the
boundary time and then decreases after meeting a maximum. Finally
it reaches a constant at the late time (Fig.\ref{l=1&l=5}). This
phenomenon is different from the evolution of entanglement
entropy, which never decreases during the whole stage of the
evolution. Moreover, the width of the strip also effects the
evolution of the complexity. When the strip is narrow, the
complexity evolves continuously while when the strip becomes
wider, it evolves discontinuously at some moment (see
Fig.\ref{l=1c} and Fig.\ref{l=5c}respectively). This discontinuity
can be understood as following: The gray line in Fig.\ref{l=5c}
corresponds to the swallow tail in Fig.\ref{l=5s}. Since we only
keep the solutions with minimum area, the system does not undergo
the evolution along the gray line but just dropping down
vertically, as shown in Fig.\ref{l=5c}.

\begin{figure}
  \centering
\subfigure[]{\label{l=5&Q=065}
\includegraphics[width=200pt]{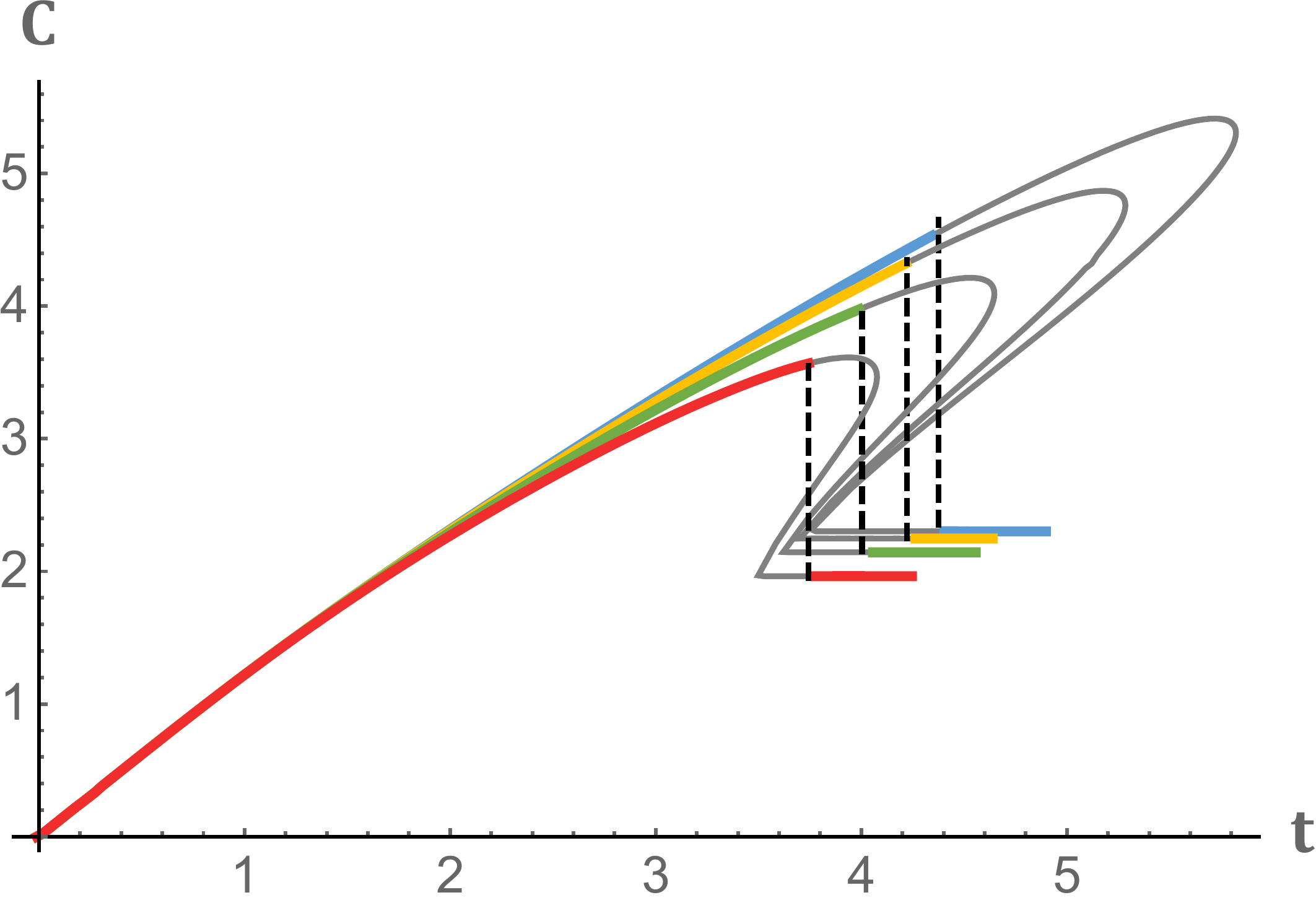}}
\hspace{40pt}
\subfigure[]{\label{l=1&Q=065}
\includegraphics[width=200pt]{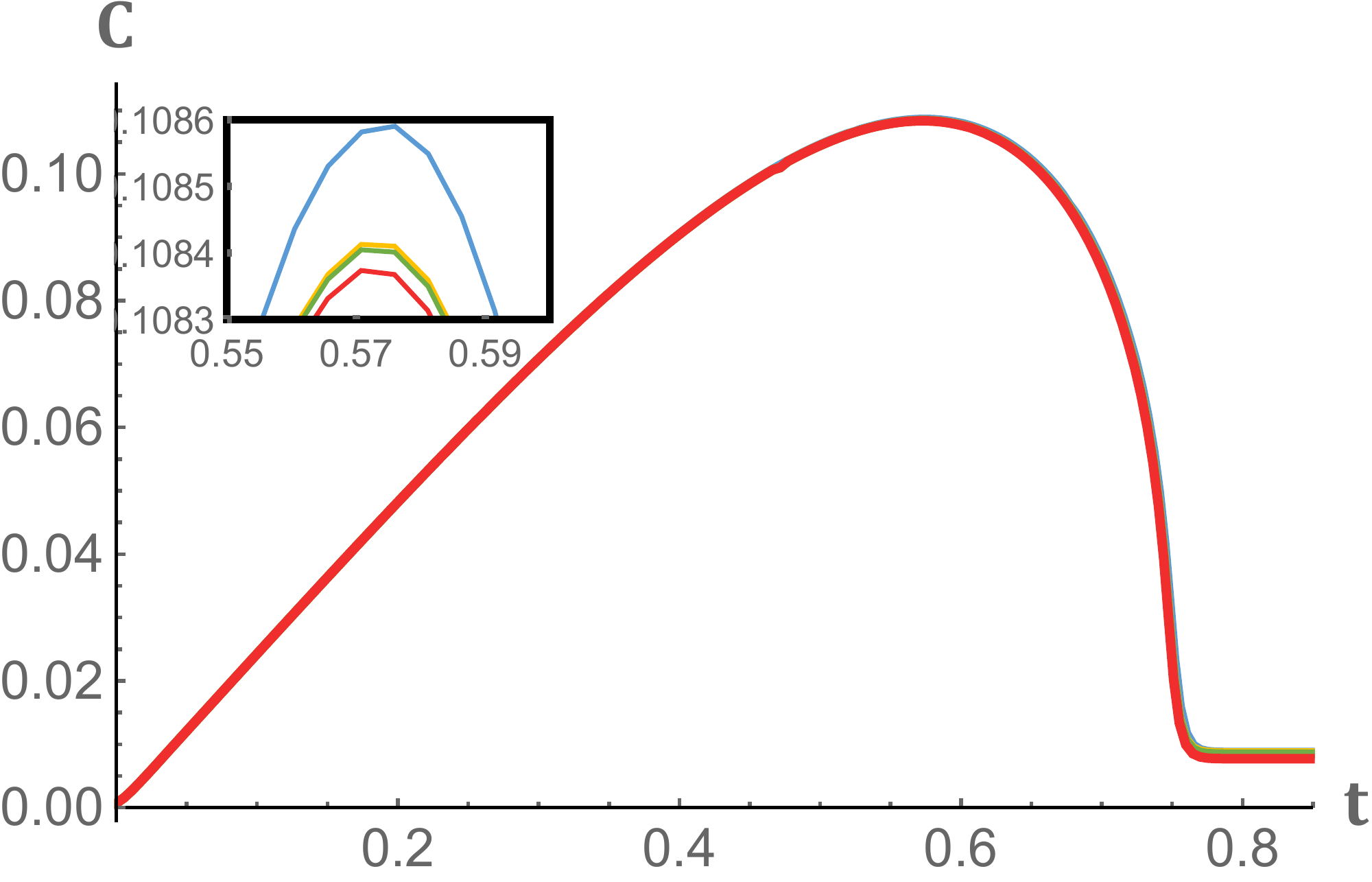}}

\caption{The dependence of the holographic subregion complexity
on the parameter $b$ with $l=5$ (the left plot) and $l=1$ (the
right plot). The blue, yellow, green and red lines correspond to
$b=0$, $\frac{1}{2}$, $1$, and $2$ respectively, with a fixed
charge $Q=0.65$.
}\label{Q}
\end{figure}

Our above result is in agreement with the previous one obtained in
the Vaidya-AdS spacetime\cite{Chen:2018mcc}. As argued in
\cite{Stanford:2014jda}, the growth of the complexity is measured
by the growth of the region inside the black hole. For a quench,
we notice that during the evolution the extremal surface
$\Gamma_{\mathcal{A}}$ stretches into the interior of the black
brane at first and then be squeezed out. Thus we tend to interpret
the above results as: during the evolution the growth of the
subregion complexity results from the fact that the extremal
surface $\Gamma_{\mathcal{A}}$ starts to probe the interior of the
black brane, while finally its dropping down at later times
reflects the fact that the surface $\Gamma_{\mathcal{A}}$ is being
squeezed out of the black brane.

In next subsection we will investigate the dependence of the
subregion complexity on the charge $Q$ and the parameter $b$, and
its distinct behavior from that of HEE will be addressed.

\begin{figure}
  \centering
\subfigure[]{\label{l=5&Q=04}
\includegraphics[width=200pt]{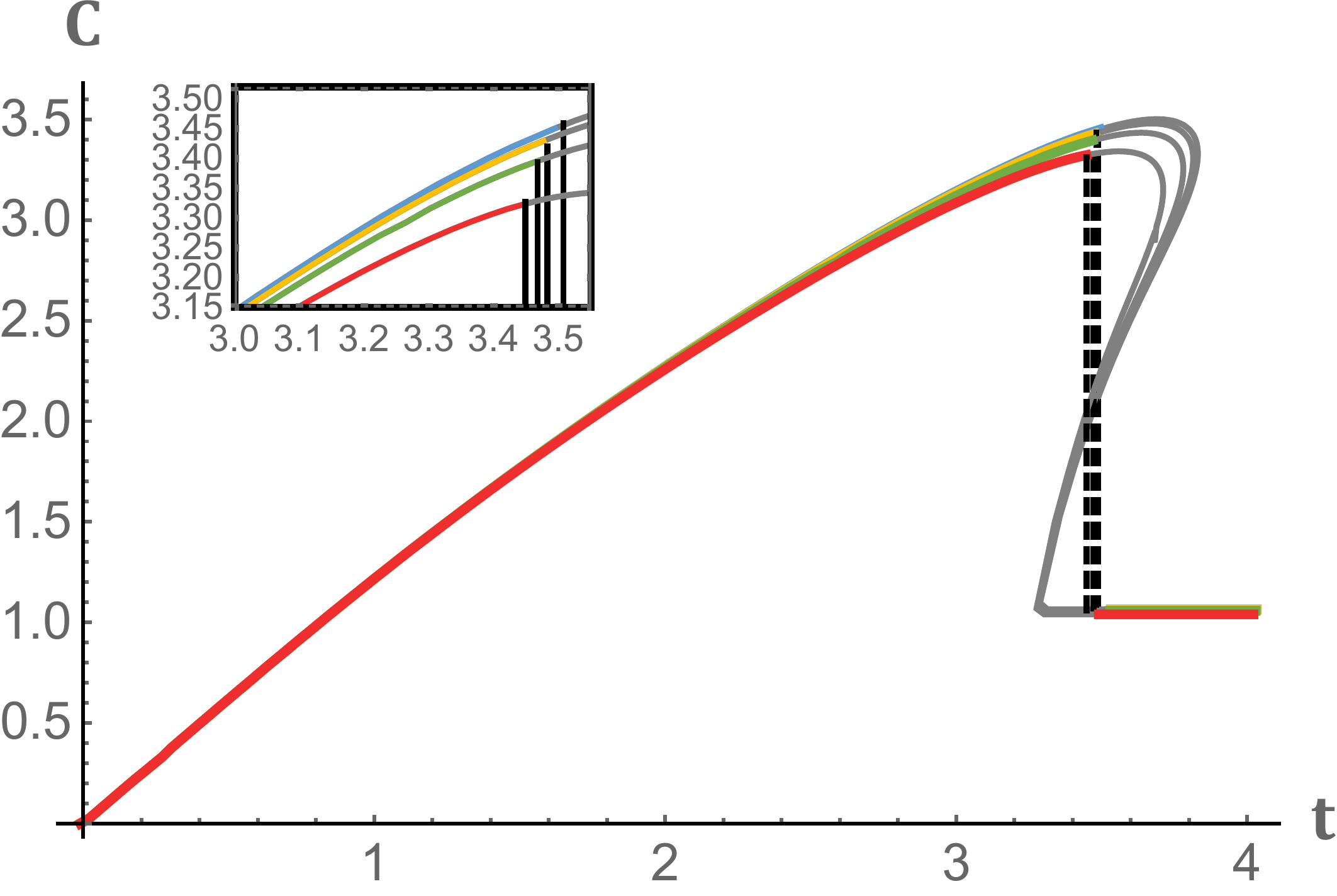}}
\hspace{40pt}
\subfigure[]{\label{l=5&Q=02}
\includegraphics[width=200pt]{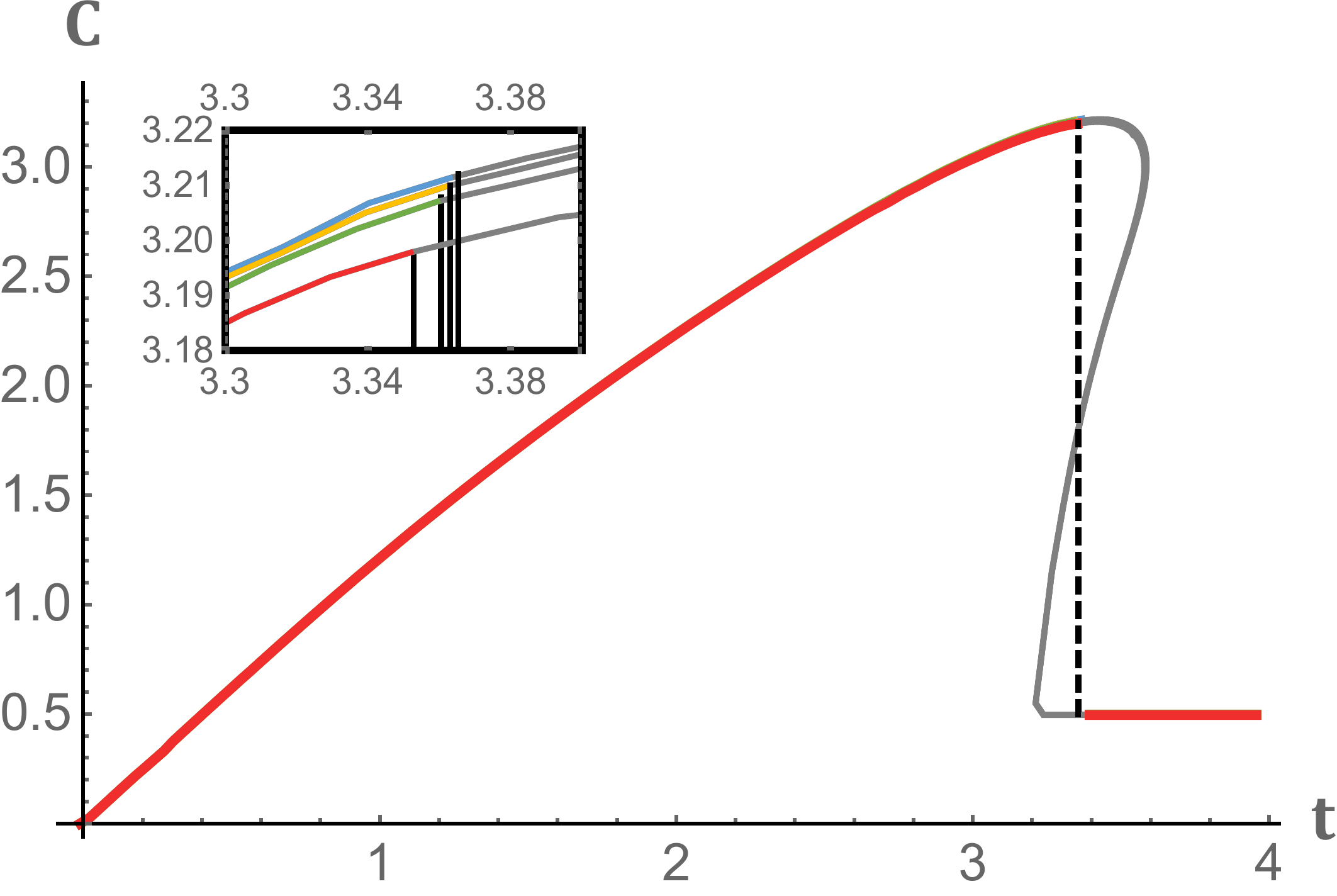}}
\caption{ The dependence of the holographic subregion
complexity on the parameter $b$ with $Q=0.4$ (the left plot) and
$Q=0.2$ (the right plot). The blue, yellow, green and red lines
correspond to $b=0$, $\frac{1}{2}$, $1$, and $2$ respectively,
with a fixed width $l=5$.
}\label{Q2}
\end{figure}

\subsubsection{Dependence on parameter $b$}

In Fig.\ref{Q} we illustrate the evolution behavior of subregion
complexity with different values of parameter $b$, while the
charge $Q$ and the width of the strip $l$ are fixed. As we can
see, at the early stage the growth rates of the complexity are
almost the same for different $b$. However, at later time the
effect of $b$ becomes important. The smaller the parameter $b$ is,
the longer the subregion complexity grows and the larger the stable value is. That is to say, the nonlinear feature of
the bulk theory prevents the subregion complexity from growing in
its dual CFT.

Another novel feature of complexity observed here is that its maximal value increases with the decrease of the parameter $b$, which is in contrast to
the behavior of the entanglement entropy. As demonstrated in
(Fig.\ref{l=5&Q=0.65&S}), while decreasing $b$, the maximal value
of HEE decreases.

Finally we remark that the discrepancy of the curves in four
colors becomes more evident in the background with large charge
$Q$ as shown in Fig.\ref{Q2}. This is reasonable since the parameter $b$ characterizes
the nonlinearity of electromagnetical field. When the value of
charge $Q$ is small, the contribution of electromagnetical field
becomes less important.

\subsubsection{Dependence on the charge $Q$}
In this subsection, we study the impact of the charge $Q$ on the
evolution of subregion complexity when the parameter $b$ is fixed.
The relevant results are plotted in Fig.\ref{l=5,b} and
Fig.\ref{l=1,b}. At the early stage, the growth rate of complexity
is almost the same for different values of charge $Q$, while at
later time the effect of charge $Q$ becomes more significant. We
find the smaller the charge $Q$ is, the sooner the subregion
complexity drops down and the smaller the maximum complexity is.
We remark that this result is in contrast to the evolution of
entanglement entropy as well, where the maximum of entanglement
entropy increases when decreasing the charge $Q$, as shown in
Fig.\ref{l=5&b=2&S}.

In both Fig.\ref{l=5,b} and Fig.\ref{l=1,b}, the stable values increase with the charge $Q$. But in the $3$-dimensional case as shown in Fig.\ref{d=2,l}, the stable values decrease with the charge $Q$.

In the 3-dimensional case when the charge $Q$ is sufficiently large, the final constant value of the holographic subregion complexity is always less than its initial value regardless the width $l$, as illustrated in Fig.\ref{d=2,l}. But in the 4-dimensional case, only when the width $l$ and the charge $Q$ are both very small, the final stable value could be less than the initial value. Actually, the appearence of negative stable value is quite common while investigating quench process (see \cite{Chen:2018mcc} and \cite{Ageev:2018nye}). This fact can be interpreted as follows: Complexity measures the ``difference'' between two states. Therefore, negative stable value means that the ``difference'' between the finial state and some reference state is less than the ``difference'' between the initial state and the reference state. In holographic scenario, since the reference state is still unclear, the appearance of negative value is acceptable.

\begin{figure} 
  \centering
\subfigure[]{\label{l=5&b=2}
\includegraphics[width=160pt]{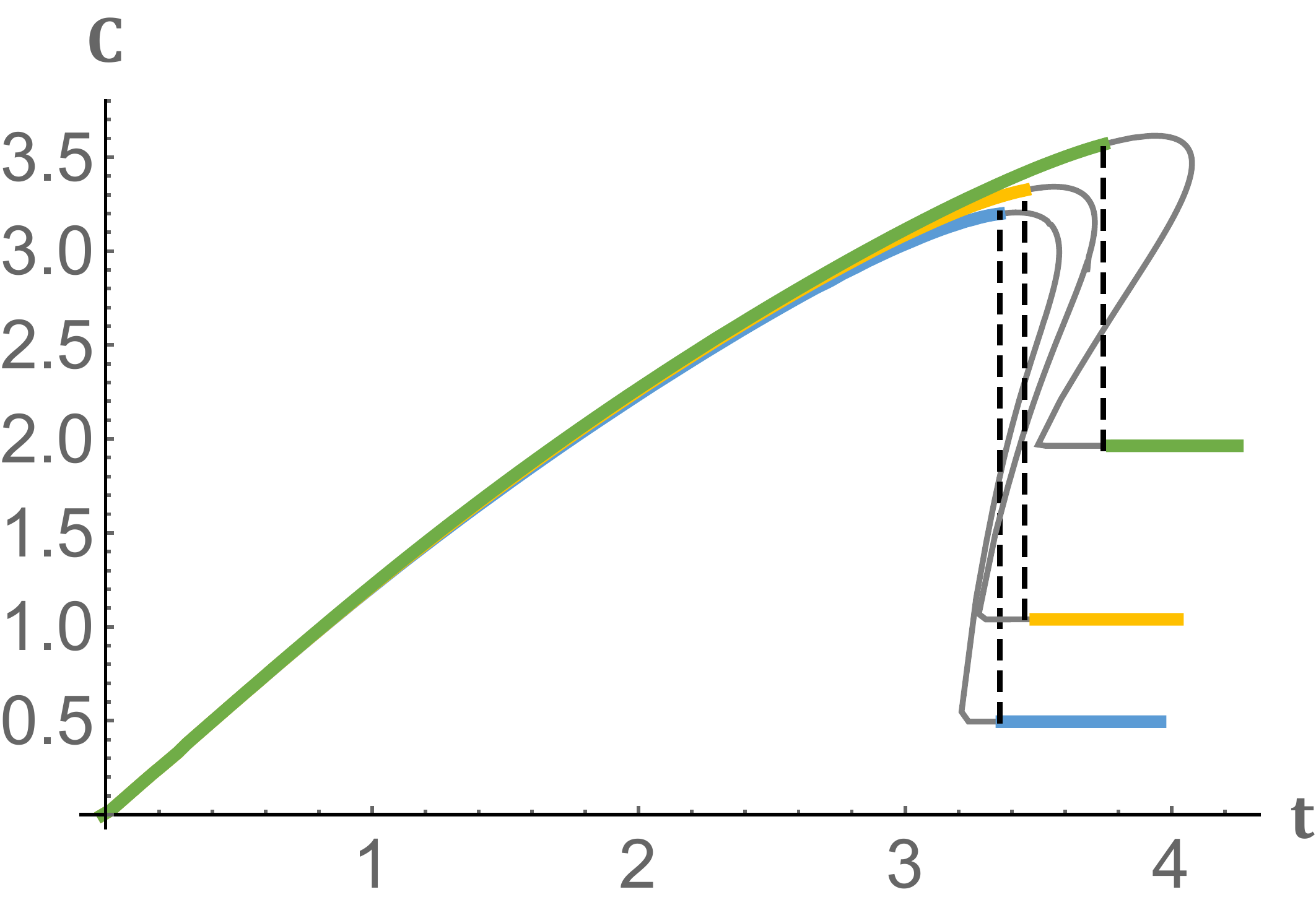}}
\hspace{0pt}
\subfigure[]{\label{l=5&b=1}
\includegraphics[width=160pt]{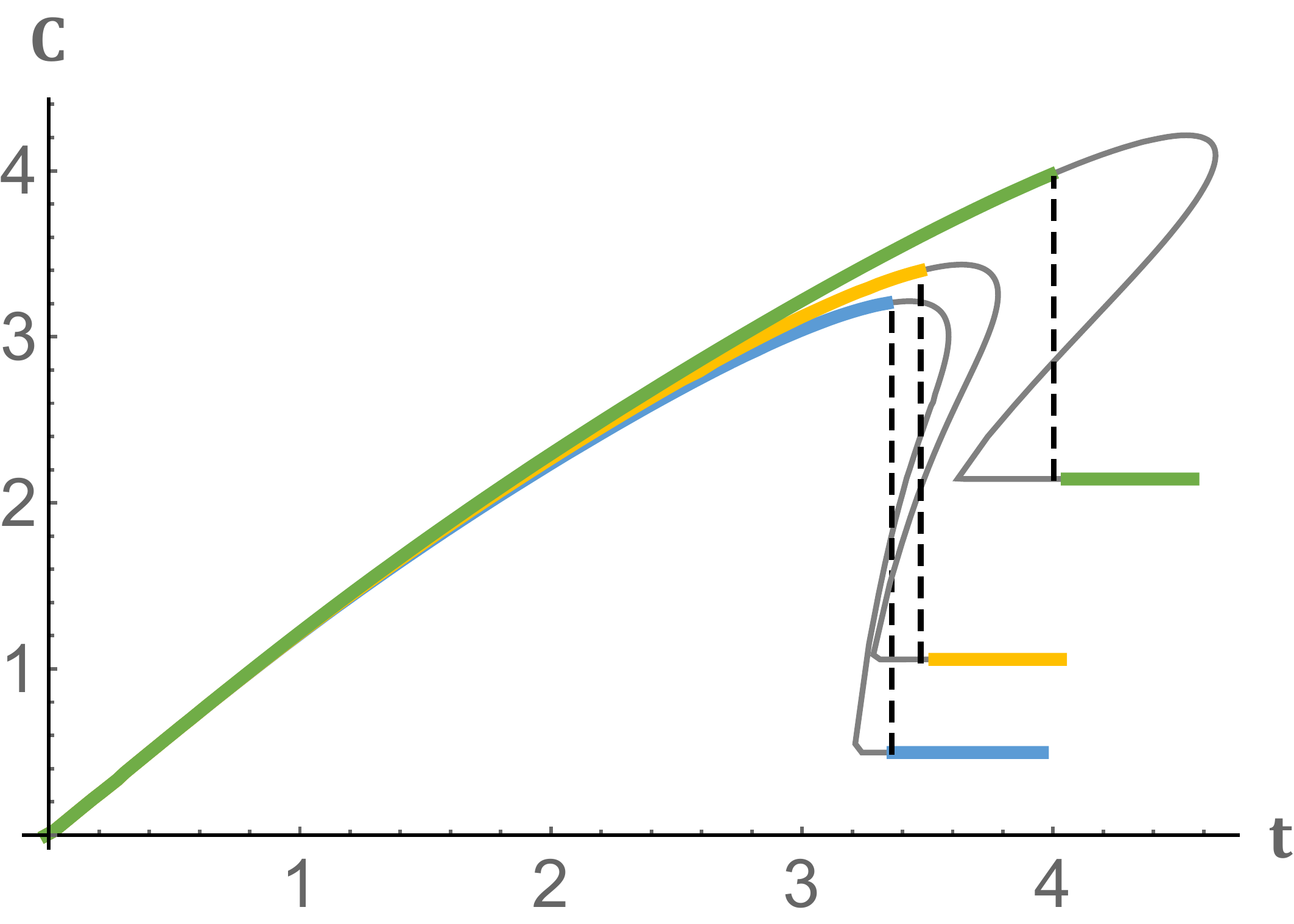}}
\hspace{0pt}
\subfigure[]{\label{l=5&b=05}
\includegraphics[width=160pt]{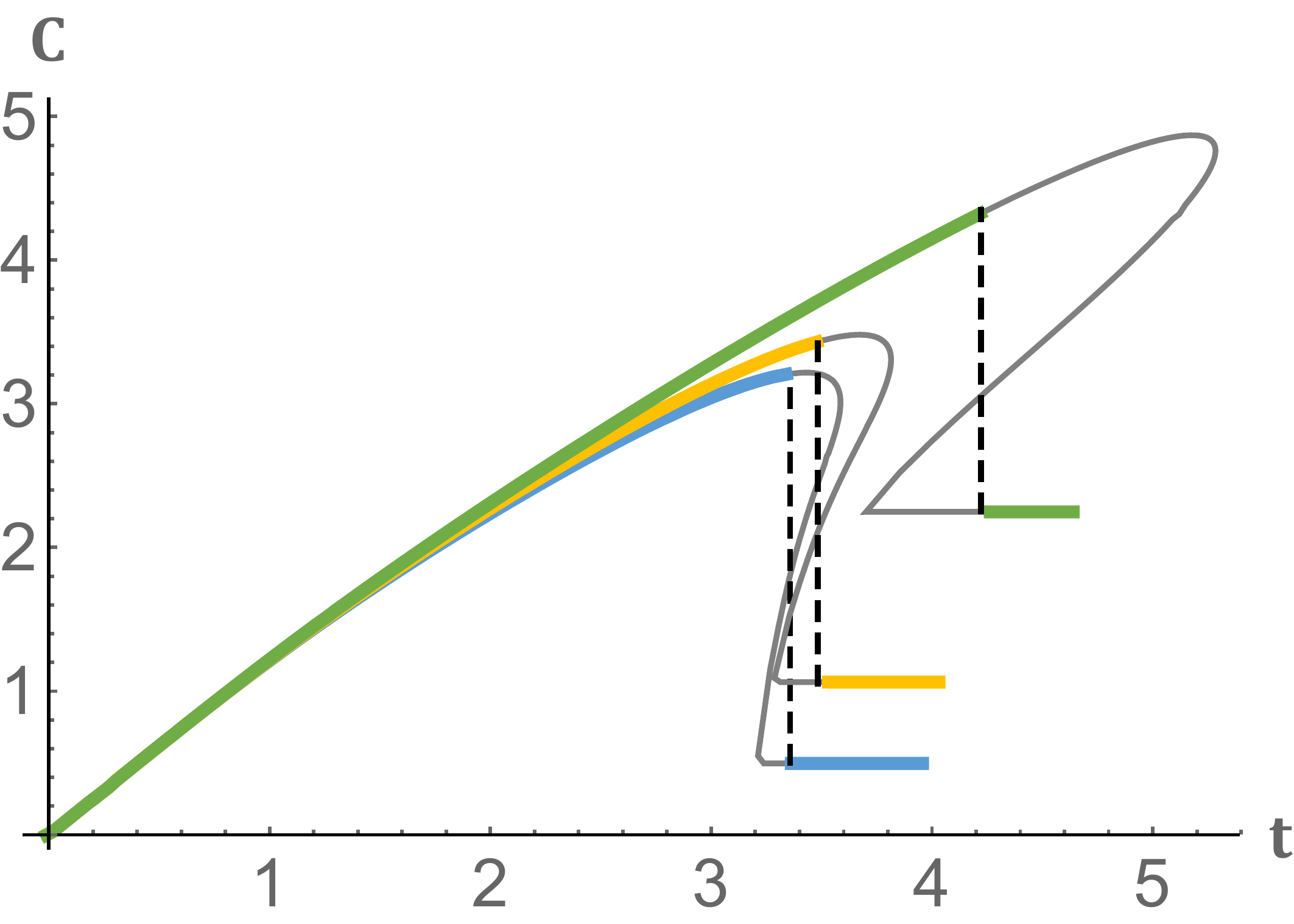}}
\hspace{0pt}
\subfigure[]{\label{l=5&b=0}
\includegraphics[width=160pt]{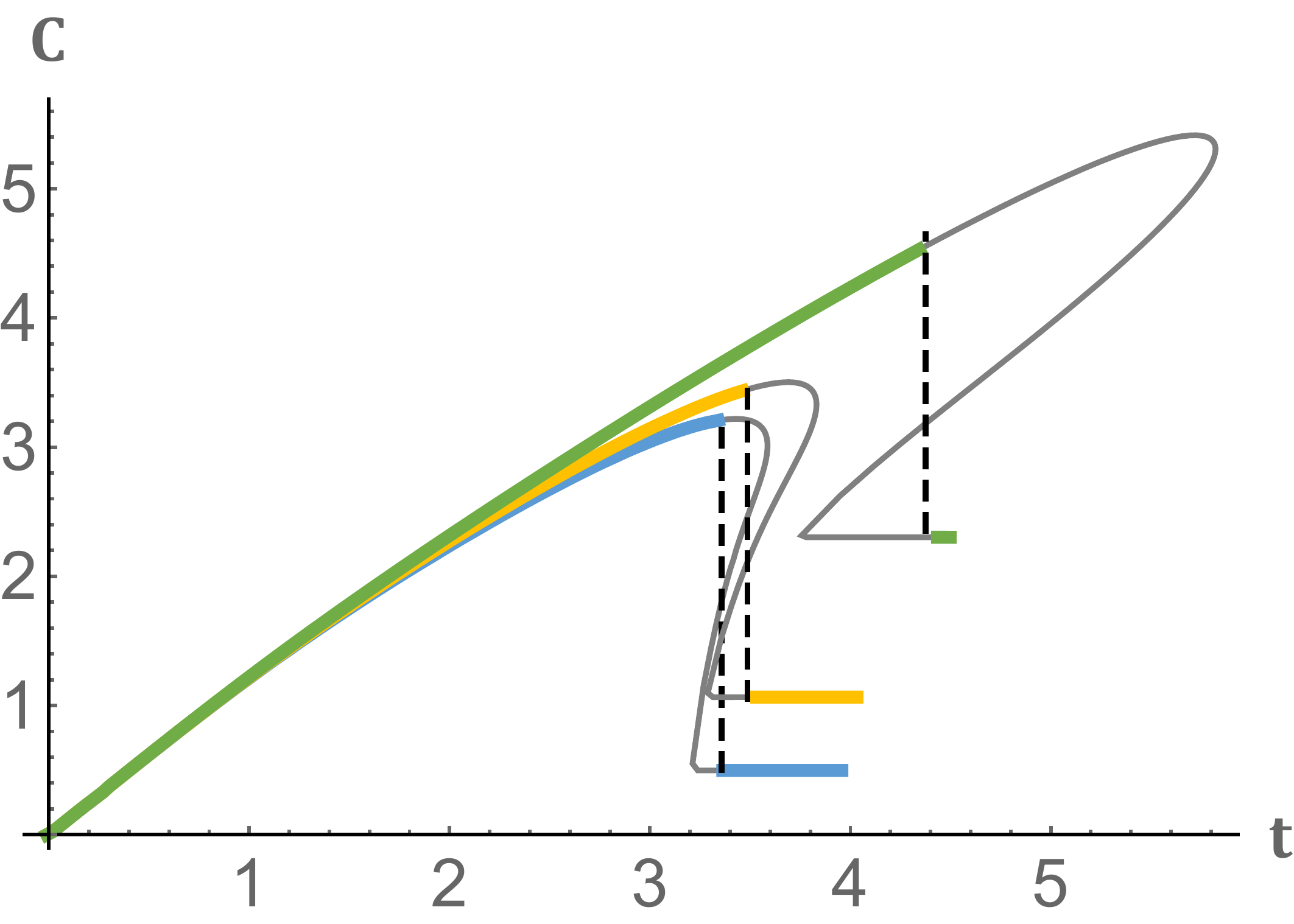}}
\caption{The dependence of the holographic subregion complexity
on the parameter $Q$ with $l=5$. The blue, yellow, green lines
correspond to $Q=0.2$, $Q=0.4$ and $Q=0.65$, respectively. The
parameter $b$ is fixed as $b=2$, $1$, $1/2$ and $0$ in subfigure
(a) to (d), respectively.
}\label{l=5,b}
\end{figure}

\begin{figure}
  \centering
\includegraphics[width=250pt]{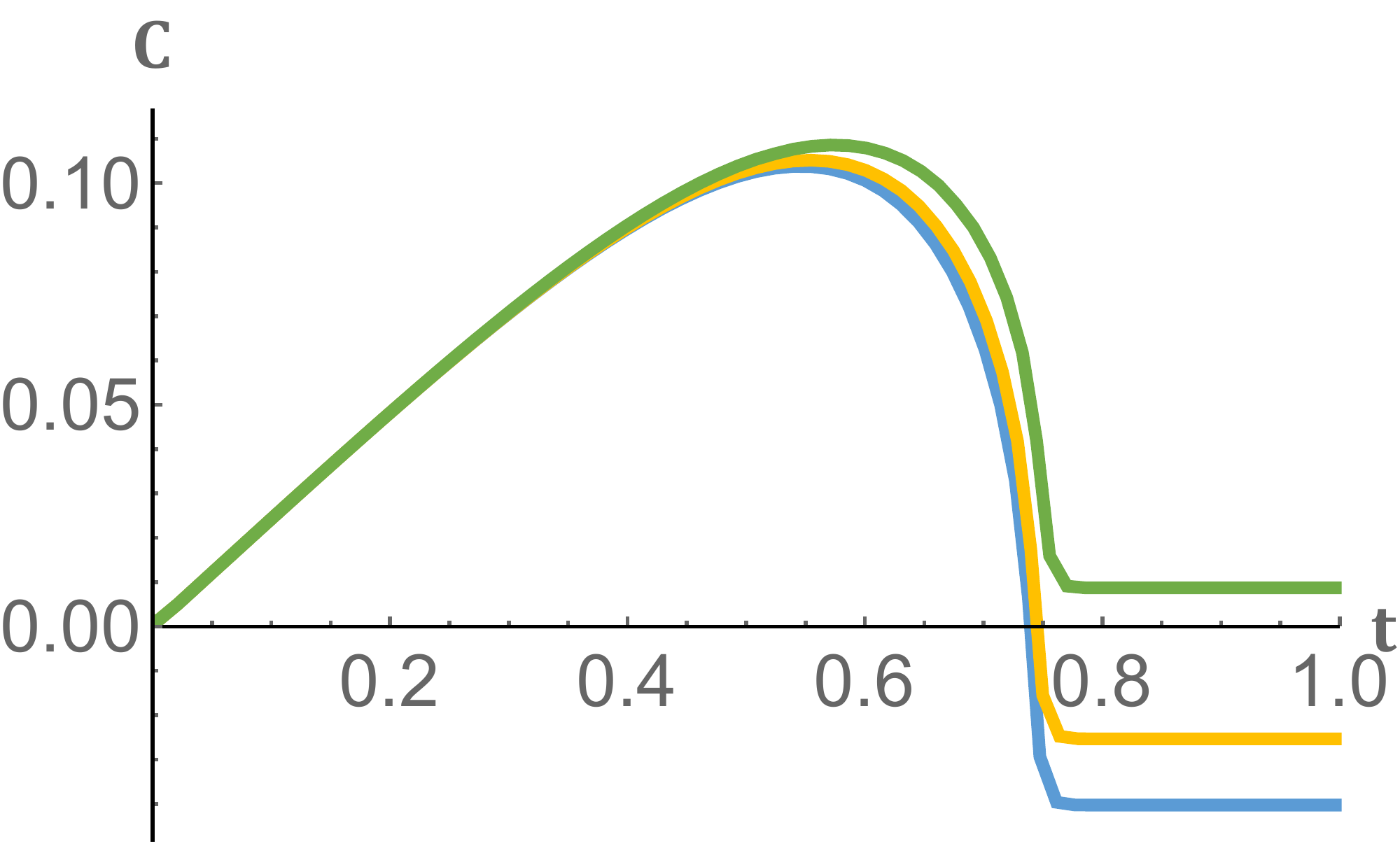}

\caption{The dependence of the holographic subregion complexity
on the parameter $Q$ with $l=1$. The blue, yellow, green lines
correspond to $Q=0.2$, $Q=0.4$ and $Q=0.65$, respectively. The
parameter $b$ is fixed as $b=0$.
}\label{l=1,b}
\end{figure}

It is interesting to compare the impacts of charge Q and parameter
$b$ on the stable value. On the one hand, we notice the effects of
the charge $Q$ is always evident, regardless the value of
parameter $b$ and width $l$, as illustrated in Fig.\ref{l=5,b} and
Fig.\ref{l=1,b}. On the other hand, in Fig.\ref{Q} we notice that
only when the width $l$ is large and the value of charge $Q$ is
close enough to its extremal value, then the effect of $b$ becomes
obvious.

In the remainder of this section we focus on the effects on charge
$Q$ on the evolution pattern of the complexity. As found in
\cite{Chen:2018mcc}, the evolution of complexity density (which
means the complexity in the unit of width $l$) in $4$-dimensional
Vaidya-AdS spacetime shows a transition from a pattern of
continuous growth into a pattern of discontinuous growth. But in
$3$-dimensional case, the evolution always exhibits a continuous
growth pattern. In addition, when the width $l$ is large enough,
the growth exhibits two distinct stages: the first rapid growth
and the second linear growth as shown in Fig.\ref{d=2,l=15}. The
above results are obtained in the neutral case\cite{Chen:2018mcc}.
Now when the black brane is charged, the Vaidya-RN-AdS metric in
$3$-dimensional spacetime can be given as \cite{Caceres:2012em}
\begin{align}
  ds^{2}=\frac{1}{z^{2}}(-f(z,v)dv^{2}-2dvdz+dx^{2}) \label{3dRNAdS} \\
  f(z,v)=1-m(v)z^{2}+q(v)^{2}z^{2}log(z)\nonumber,
\end{align}
where $m(v)$ and $q(v)$ are shown in equation
(\ref{eq:Massfunction}).

\begin{figure}
  \centering
\subfigure[]{\label{l=5&d=2}
\includegraphics[width=200pt]{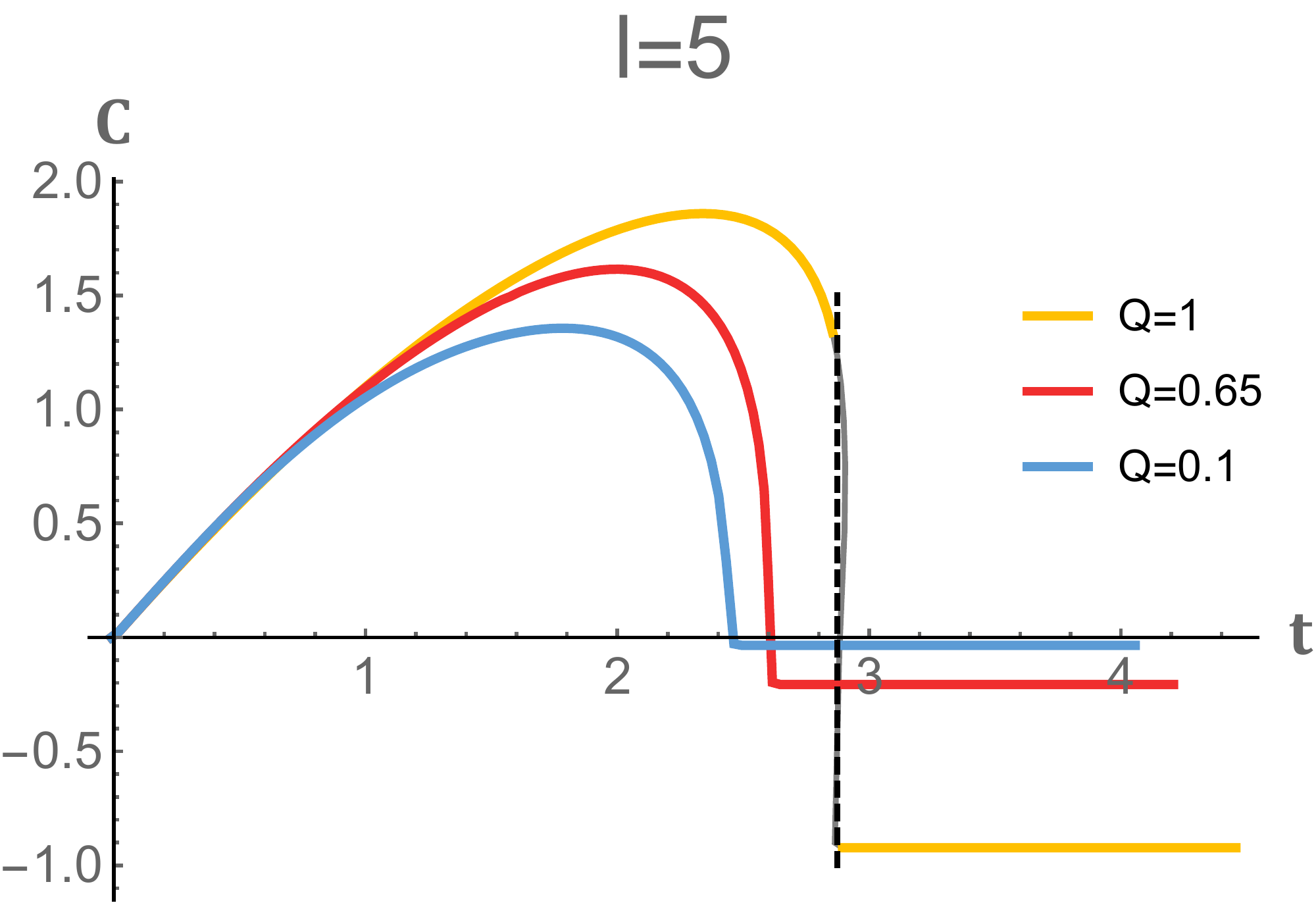}}
\hspace{0pt}
\subfigure[]{\label{d=2,l=15}
\includegraphics[width=200pt]{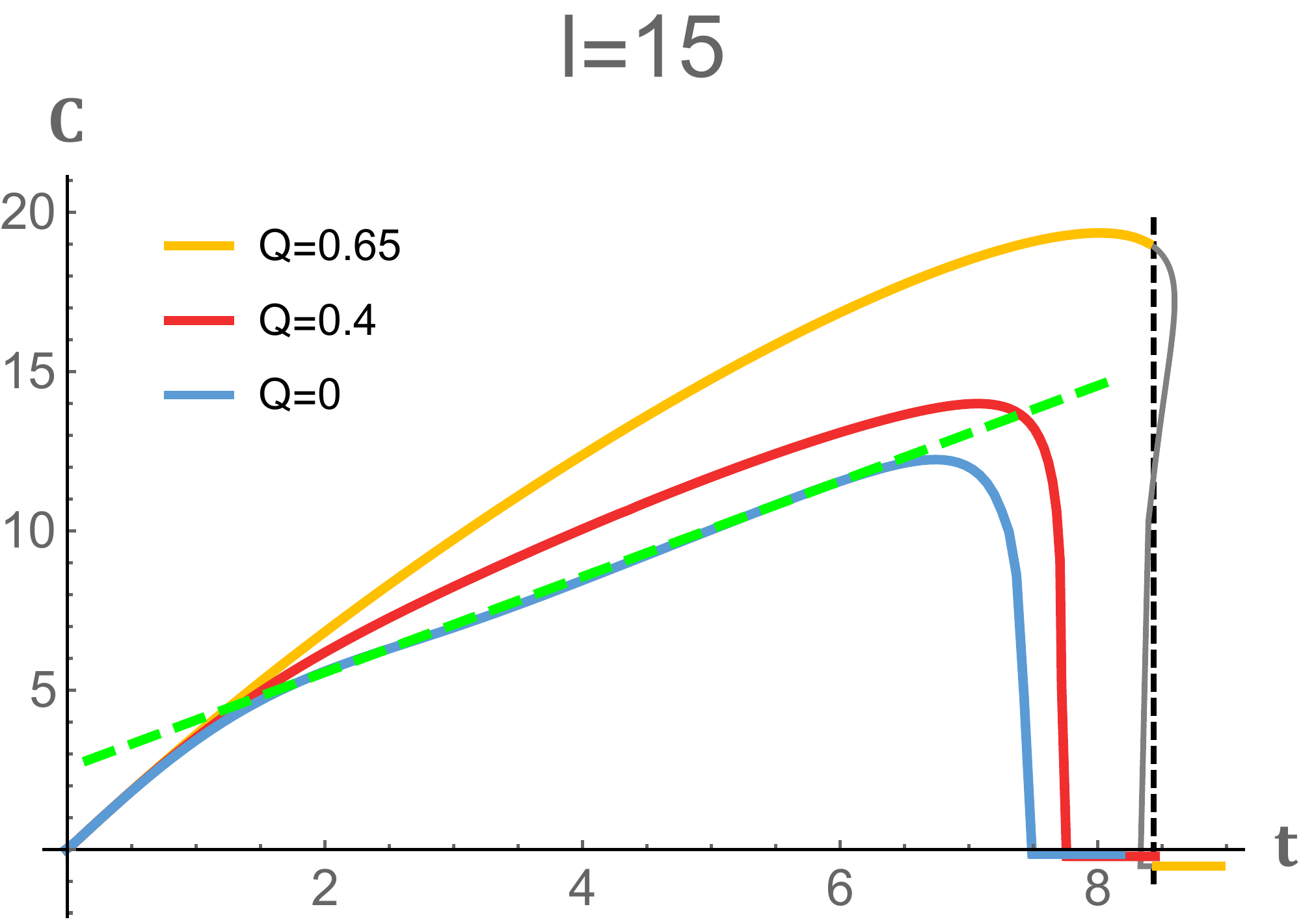}}

\caption{The impact of the charge $Q$ on the evolution of
subregion complexity in three dimensional RN-AdS background
($b=0$). In Fig.\ref{d=2,l=15}, the green dashed line
represents the linear growth stage at later time when $l$ is large
enough. }\label{d=2,l}
\end{figure}

It is quite straightforward to obtain the complexity for charged
black branes, as plotted in Fig.\ref{l=5&d=2} and
Fig.\ref{d=2,l=15}. Interestingly enough, we find the charge $Q$
can not only change the growth behavior, but also change the
pattern of evolution. When the charge $Q$ is large enough, the
evolution of complexity changes the pattern from continuous to
discontinuous (Fig.\ref{d=2,l}). Moreover, the sufficiently large
charge $Q$ will wash out two different growth stages. This result
can be read from Fig.\ref{d=2,l=15}, where the blue line
represents the case of AdS-Schwarzschild background and we can see
two distinct growth stage clearly. That is to say, with
sufficiently large charge $Q$ the evolution of complexity shows a
transition from the continuous pattern into the discontinuous
pattern and forgets about its later linear growth stage (Fig.
\ref{l=5&d=2} and \ref{d=2,l=15}).

\section{Conclusions and discussions}

In this paper we have investigated the evolution of the subregion
complexity during a quench in Einstein-Born-Infeld theory. The
subregion $\mathcal{A}$ we consider here is an infinite strip on a
time slice of the boundary. Holographically the subregion
complexity can be described by a codimension-one extremal
hypersurface $\Gamma_{\mathcal{A}}$ in the bulk. We have
numerically analyzed the evolution behavior of holographic
entanglement entropy and the subregion complexity, which
geometrically reflect the evolution of the HRT surface and the
codimension-one extremal hypersurface during the course of the
quench. The increasing and decreasing behavior of the subregion
complexity are related to the part which is stretched into the
black brane. we have also investigated the effect of varying the
charge $Q$ and the parameter $b$ on the evolution of the
complexity. It turns out that the maximum of the complexity drops
down when we decrease the charge $Q$ or increase the parameter
$b$. Moreover, when the charge $Q$ is large enough, it washes out
the second stage featured by linear growth. But under the limit of
$l\rightarrow \infty$, we tend to interpret this effect as
retarding the occurrence of the second stage of linear growth
rather than washing out it directly. One should be cautious to
extend this result to the limit of $l\rightarrow \infty$, since
the width of the strip $l$ in numerical simulation perhaps is not
large enough to probe the whole region due to the numerical
limitation. When the charge $Q$ is sufficiently large, whether the
linearly growing stage would appear should be tested analytically
with the strategy as proposed in \cite{Liu:2013qca}. And more
detail of these results should be explored in an analytical way
too. In addition, these results should be helpful for us to
further disclose the role of subregion complexity in the direction
of understanding the holographic nature of space time.

It should be interesting to explore the evolution of complexity
analytically under the subregion CV or CA conjecture. It is also
desirable to investigate the min flow-max cut theorem in the
Vaidya-type spacetime to build the quantum gates in the bulk.
Further, we should note that the features which can be probed by
the holographic subregion complexity is also sensitive to the HEE
in this paper. It is quite intriguing to investigate the
evolution behavior of the complexity in the circumstance that is
insensitive to the HEE in future.

\section*{Acknowledgments}

We are very grateful to Chao Niu and Runqiu Yang for helpful
discussions and suggestions. This work is supported by the Natural
Science Foundation of China under Grant No. 11575195 and
11875053. Y.L. also acknowledges the support from Jiangxi young
scientists (JingGang Star) program and 555 talent project of
Jiangxi Province. C.-Y. Zhang is supported by National Postdoctoral Program for Innovative Talents BX201600005 and Project funded by China Postdoctoral Science Foundation.

\end{document}